\documentclass[aps,prl,preprint,superscriptaddress,floatfix]{revtex4-1}
\usepackage[T1]{fontenc}
\usepackage{graphicx}% Include figure files
\usepackage{dcolumn}% Align table columns on decimal point
\usepackage{bm}% bold math
\usepackage{color}

\begin{document}

\title[Activated wetting of nanostructured surfaces: reaction coordinates,
finite size effects, and simulation pitfalls]{Activated wetting of
nanostructured surfaces: reaction coordinates, finite size effects, and
simulation pitfalls}

\author{M. Amabili}
\email{matteo.amabili@uniroma1.it}
\affiliation{Dipartimento di Ingegneria Meccanica e Aerospaziale, Universit\`a
di Roma ''La Sapienza'', Rome, Italy}
\author{S. Meloni}
\affiliation{Dipartimento di Ingegneria Meccanica e Aerospaziale, Universit\`a
di Roma ''La Sapienza'', Rome, Italy}
\author{A. Giacomello}
\affiliation{Dipartimento di Ingegneria Meccanica e Aerospaziale, Universit\`a
di Roma ''La Sapienza'', Rome, Italy}
\author{C.M. Casciola}
\affiliation{Dipartimento di Ingegneria Meccanica e Aerospaziale, Universit\`a
di Roma ''La Sapienza'', Rome, Italy}


%\begin{tocentry}
%\includegraphics[width=1.0\textwidth]{toc.pdf}
%\end{tocentry}

\begin{abstract}
A liquid in contact with a textured surface can be found in two states, Wenzel
and Cassie. In the Wenzel state the liquid completely wets the corrugations
while in the Cassie state the liquid is suspended over the corrugations with
air or vapor trapped below. The superhydrophobic properties of the Cassie
state are exploited for self-cleaning, drag reduction, drug delivery etc.,
while in the Wenzel state most of these properties are lost; it is therefore of
great fundamental and technological interest to investigate the kinetics and
mechanism of the Cassie-Wenzel transition. Computationally, the Cassie-Wenzel
transition is often investigated using enhanced sampling (``rare events'')
techniques based on the use of collective variables (CVs). The choice of the
CVs is a crucial task because it affects the free-energy profile, the
estimation of the free-energy barriers, and the evaluation of the mechanism of
the process. Here we investigate possible simulation artifacts introduced by
common CVs adopted for the study of the Cassie-Wenzel transition: the average
particle density in the corrugation of a textured surface and the coarse
grained density field at various levels of coarse graining. We also investigate
possible additional artifacts associated to finite size effects. We focus on a
\emph{pillared} surface, a system often used in technological applications. We
show that the use of the average density of fluid in the interpillar region
brings to severe artifacts in the relative Cassie-Wenzel stability and in the
transition barrier, with an error on the estimates of the free energies of up
to hundreds of $k_BT$ and wrong wetting mechanisms. A proper description of the
wetting mechanism and its energetics requires a rather fine discretization of the density
field. Concerning the finite size effects, we have found that the typical
system employed in the Cassie-Wenzel transition, containing a single pillar
within periodic boundary conditions, prevents the break of translational
symmetry of the liquid-vapor meniscus during the process. Capturing this break of
symmetry is crucial for describing the transition state along the wetting
process, and the early stage of the opposite process, the Wenzel-Cassie
transition. 

%Valid PACS numbers may be entered using the \verb+\pacs{#1}+ command.
\end{abstract}

%\pacs{05.20Jj, 68.03.Cd, 68.08.Bc}
%\maketitle

\maketitle

\section{Introduction}
\label{sec:intro}   
  
A liquid in contact with a submerged nanotextured surface can be either in the
Cassie \cite{cassie1944} or  Wenzel \cite{wenzel1936} state. In the Cassie
state, a vapor or gas layer sustains the liquid, which is, thus, in contact
only with the top of the surface textures. The presence of a composite
vapor-liquid-solid interface makes the surface superhydrophobic, i.e.
characterized by properties such as enhanced slippage \cite{rothstein2010} or
self-cleaning.~\cite{bhushan2009} In the Wenzel state, instead, the liquid
completely fills the nanotextures with the ensuing loss of all the
superhydrophobic properties.  

In normal conditions the transition from Cassie to Wenzel, or the
reverse dewetting transition, is a \emph{thermally activated}
(stochastic) process, i.e.  the system must overcome a free energy
barrier in order to go from one state to the other. According to the
transition state theory~\cite{eyring1935}, the height of the free energy
barrier determines the kinetics of the process.  Indeed, the (average)
time needed to observe a transition, or equivalently the lifetime of a
given state, depends exponentially on the ratio between the values of
the barrier and of the thermal energy $k_B T$, with $k_B$ the Boltzmann
constant and $T$ the temperature.  The presence of large free-energy
barriers has been confirmed experimentally. Indeed, very long Cassie
state lifetimes, of the order of tens of minutes to tens of hours, have
been observed for textured surfaces subject to moderate liquid
pressures ($\approx 5\%$ higher than the atmospheric
pressure~\cite{xu2014,bobji2009}).

Molecular dynamics, thanks to its minimal assumptions on the model of the
three-phase system (solid, liquid, vapor), seems an optimal tool to investigate
the mechanism and kinetics of the Cassie-Wenzel transition on nanotextured
systems, and their dependence on the surface topography and on the
thermodynamic conditions. However, due to the long transition time connected
with large free energy barriers, special simulation techniques are needed to
investigate the collapse of the superhydrophobic state. Thus, the Cassie-Wenzel
transition has been studied using several advanced sampling techniques: boxed
molecular dynamics, \cite{savoy2012b} forward flux sampling, \cite{savoy2012a}
restrained molecular dynamics, \cite{giacomello2012,amabili2016} (indirect)
umbrella sampling, \cite{prakash2016} and the string
method.~\cite{giacomello2015} All these methods require the use of one or more
(typically few) collective variables (CVs), i.e., a set of observables which
can describe the macroscopic configuration of the system along the transition.
In most of the simulations a single CV has been used, namely the number of
fluid particles within a control volume enclosing the surface corrugations,
$n_l$.~\cite{giacomello2012,savoy2012a,amabili2016,prakash2016} In a two-phase
liquid-vapor system this CV is related to the amount of liquid in the same
region. Since in the Cassie state the liquid is pinned at the top of the
texture while in the Wenzel state it wets the corrugations completely, $n_l$
seems the most natural variable to characterize the process. $n_l$ has several
advantages: i) it is relatively simple to use in atomistic simulations; ii) it
is an atomistic equivalent of the volume of liquid in the corrugations, which
allows to establish a connection and compare results with the continuum theory
of wetting of confined systems, CREaM~\cite{giacomello2012a}; iii) considering
the reverse process, dewetting, it is related to the liquid density and allows
to establish connections with the Lum-Chandler-Weeks's theory of

hydrophobicity~\cite{lum1999} (see Refs.~\citenum{remsing2015,amabili2016}).
However, for the specific case of the 2D square cavity it has been shown that
the (coarse-grained) density field is necessary to properly describe the
wetting transition and accurately determine the corresponding
barrier.~\cite{giacomello2015} Despite the conclusions of this latter work, and
the  known risk that an unfortunate choice of CVs can critically affect
simulation results,~\cite{bolhuis2002} a limited attention has been devoted to
identify possible artifacts introduced by the use of $n_l$. This is especially
important in the case of surfaces with 3D interconnected cavities, such as
pillared structures, for which it has been reported that the transition takes
place with changes in the morphology of the liquid-vapor
meniscus.~\cite{prakash2016} 

Another aspect that might introduce artifacts in the study of the wetting
transition is the size of the simulation box. In the case of artificial
textured surfaces, this is translated into the number of \emph{unit cells}
which are included in the computational sample, e.g., the number of pillars for
a pillared surface. For example, pillared superhydrophobic surfaces have been
typically modeled using a simulation box containing a single unit cell
consisting of the repeated motif composing the actual system: one pillar and
the associated part of the bottom wall (see
Fig.~\ref{fig:geometria}).~\cite{hemeda2014,zhang2014,prakash2016,tretyakov2016,panter2016}
This simulation setup prevents the breaking of translational symmetry of the
liquid meniscus in the directions of the surface during wetting, which have
been observed in a number of recent
experiments.~\cite{butt2013,lv2014b} This limitation is especially important
because symmetry breaking, and its origin, whether intrinsic of Cassie-Wenzel
transitions~\cite{giacomello2012b} or due to the presence of dirt on the
surface,~\cite{DuanReview2016} is debated in the literature.

The aim of this work is to analyze in detail the effect of the choice of CVs
and the simulation box size on the mechanism and energetics of collapse of
superhydrophobicity in 3D textured surfaces with interconnected cavities,
namely on pillared surfaces. We focus on this type of surfaces because they are
largely used in technological applications \cite{verho2012} and in experiments,
\cite{checco2014,butt2013} are simpler than other 3D surfaces to investigate,
and yet preserve the main aspects of a large class of surface textures, namely
that the surface cavities are interconnected, which makes this research of
general interest.  Moreover, the present results could be of interest also for
other fields in which the density field is relevant to describe thermally
activated events: nucleation of crystals,
\cite{Kelton:2010fs,page2006,page2009} protein folding, \cite{miller2007}
membrane poration, \cite{fuhrmans2015,smirnova2015} etc. In these fields there
is an interest in developing coarse-grained models, which reduce the
computational cost while capturing the relevant physics. A more detailed
analysis of the physics of wetting and dewetting on nanopillared surfaces is
contained in a related article.~\cite{amabiliPRF}

Concerning the CVs, we consider the (coarse-grained) density field at various
levels of coarsening. The reason to focus on this CVs is manifold; first of all
it has been shown to be adequate to properly describe the wetting path in
atomistic simulations of simpler systems.~\cite{giacomello2015} Secondly, the
density field is the main variable of the sharp interface model and of the
(classical) density functional theory, two continuum theories which have been
successful at describing  wetting/dewetting in confined
systems~\cite{talanquer1996,Talanquer:2001,Lutsko:2008} (see also
Refs~\citenum{Kelton:2010fs,Meloni:2016it}). Finally, by changing the level
of coarse graining of the density field one can explore the two opposite
limits, the highly coarse-grained case of the average fluid density in the
surface corrugations, equivalent to the simple CV used in
Refs.~\citenum{giacomello2012,savoy2012a,amabili2016,prakash2016}, and the
case of the density field computed on a grid with spacing of few atomistic
radii or for continuum liquids.~\cite{ren2014,panter2016,pashos2016}

To compute the wetting  path, i.e., the sequence of coarse-grained density
fields (the CVs set) that the system takes on along the Cassie-Wenzel
transition, we use the zero temperature string method.
~\footnote{In Ref.~\citenum{giacomello2015} we have discussed the
importance of finite temperature effects in the wetting mechanism. In
particular, we focused on different wetting paths and on the overlap of the
related reaction tubes, i.e., the region of CV space where most (a prescribed
fraction) of the reactive trajectories pass through. While this effect could be
important also for pillared systems, the methodological aspects investigated in
this work do not depend on this type of effects. In fact, their discussion
would increase the complexity of the work without shedding new light on the
choice of CVs discussed here.}$^,$~\cite{maragliano2006,weinan2002} The
string method allows one also to determine the energetics along the transition
path.  In the case of a single CV, the average density of fluid particles in
the corrugations,  the string method reduces to a simpler techniques, the
restrained MD method (RMD~\cite{TAMD}).  As we will discuss more in detail
below, results obtained by RMD (or, equivalently, by temperature accelerated MD
-- TAMD \cite{TAMD}) are analogous to those that can be obtained with umbrella
sampling (US~\cite{TorrieValleau}), including its \emph{indirect} version
(INDUS~\cite{Patel:2010cw}), and boxed molecular dynamics
(BXD~\cite{Glowacki:2009hqa}): simulation artifacts are not due to the
specific simulation method employed but to the choice of CVs. Thus, the
conclusions  we will draw on RMD have a  broader scope: they do not refer to a
specific simulation technique but to the use of suitable or inadequate CVs.

Concerning the system, we considered  two cases: one comprising a surface with
$3 \times 3$ square pillars and one with a single pillar, i.e. only one unit
cell (Fig.~\ref{fig:geometria}).  In both cases periodic boundary conditions
are applied in the direction of the surface, $x$ and $y$. The second case is
analogous to the one adopted in
Refs.~\citenum{prakash2016,zhang2014,panter2016}. 

Summarizing, in this work we investigate possible artifacts arising from the
choice of the CVs and from the size of the computational sample, and the
correlations between these two simulation parameters. The systems investigated
are shown in Fig.~\ref{fig:geometria}. 

Understanding the effect of the coarse graining and system size on
simulation results, whether or not they introduce any sizable artifact,
is crucial to assess theories on the wetting and recovery mechanism and
energetics and to provide a practical guideline for future simulations
of this process on even more complex surface topographies. This will
help moving forward toward the \emph{in silico} design of
superhydrophobic surfaces with tailored properties. Such properties
indeed depend on the surface characteristics (topography and chemistry)
via the wetting/dewetting path, which should therefore be captured
accurately in simulations.

The article is organized as follows. In Sec.~\ref{sec:model} the atomistic
model, the string method, and the restrained molecular dynamics method are
introduced. In Sec.~\ref{sec:results} the results are presented and discussed.
Finally, Sec.~\ref{sec:conclusioni} is left for conclusions.

\section{Theory} 
\label{sec:model}
\subsection{Atomistic model of the composite system}

\begin{figure*}%[h!]
        \centering
        \includegraphics[width=1.0\textwidth]{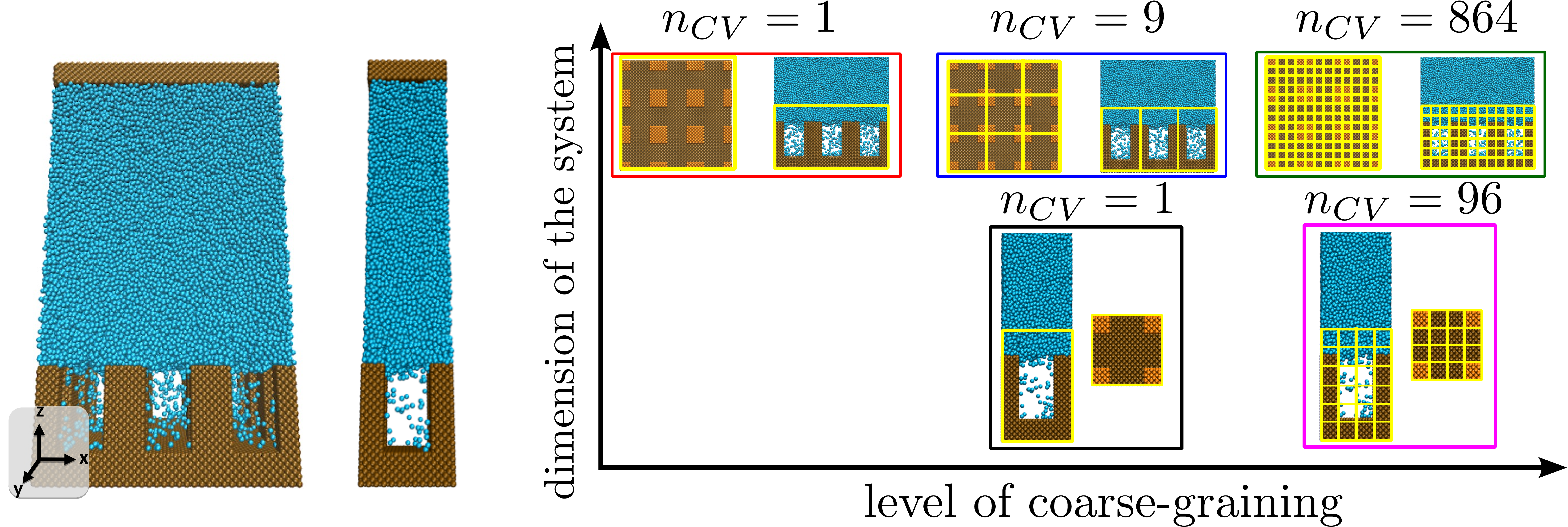}
        \caption{(Left) Atomistic systems formed by $3 \times 3$ and $1
					\times 1$ pillars. (Right)
Discretization of the space corresponding to the different sets of CVs for the
$3 \times 3$ ($1$, $9$, and $864$ CVs) and $1 \times 1$  ($1$ and $96$ CVs)
pillars systems. The coarse-grained density field corresponds to the collection
of local densities computed within the yellow parallelepipeds shown in the
figure which have size $\delta x \times \delta y \times \delta z \approx 56
\times 56  \times 29 \sigma$, $\approx 18.7 \times 18.7 \times 29
\sigma$, and $\approx 4.6 \times 4.6 \times 4.8 \sigma$ for the $1$, $9$, and
$864$ CV cases, respectively\, for the $3 \times 3$ pillars system. 
We remark that the $9$ and $864$ CV cases for the  $3 \times 3$
pillars correspond, in terms of discretization of the space ($\delta x \times
\delta y \times \delta z$), to the $1$ and $96$ CVs for the  $1 \times
1$, respectively. Comparison of results between $1 \times 1$ and $3
\times 3$ systems must be restricted to CVs corresponding to equivalent
discretizations of the density field.}
        \label{fig:geometria}
\end{figure*}

The atomistic model used to investigate the Cassie-Wenzel transition
consists of a Lennard-Jones (LJ) fluid in contact with a nanopillared LJ
solid surface (see Fig.~\ref{fig:geometria}, left panel). The fluid is
also in contact with a second smooth solid surface of LJ particles at
the top of the simulation box, which is used to control the  liquid
pressure $P_{l}$ (see below for details). Fluid and solid particles interact via a modified LJ potential

\begin{equation}
V(r_{ij})= 4 \epsilon \Bigg( \bigg(\frac{\sigma}{r_{ij}}\bigg)^{12} - c \bigg(\frac{\sigma}{r_{ij}}\bigg)^{6}\Bigg)
\label{eq:LJ}
\end{equation}
where $\sigma$ and $\epsilon$ are the characteristic length and energy of the
potential, respectively. $r_{ij}$ is the interatomic distance between particles
$i$ and $j$.  $c$ is the parameter controlling the
hydrophobicity~\cite{cottin2003} of the solid material; here we use $c=0.6$, to
which corresponds a Young contact angle~\footnote{The Young contact angle is
the angle formed between a flat surface and the tangent to a sessile liquid
droplet deposited on it at the liquid/solid contact point.} $\theta_Y \approx
113^\circ$.~\cite{giacomello2012,amabili2015} 

The fluid particles are kept at constant temperature $T=0.8\epsilon$ via
a Nos\'e-Hoover chain thermostat.~\cite{martyna1992} The pressure of the
liquid $P_{l}=0$ is controlled through the application of a constant
force $F_z$ on the particles of the upper wall, which acts as a piston,
while the bottom wall is kept fixed during the
simulation.~\cite{amabili2016} Within the quasi-static simulation
approach used in the present work, the vapor pressure $P_{v}$ depends,
essentially, only on $T$.~\cite{amabili2016Yeomans} Thus, the choice of
$F_z$ and $T$ determine the driving force of the wetting process,
$\Delta P = P_{l}-P_{v}$, which is  $\approx 0$ in the present case. 

The two investigated  systems are shown in Fig.~\ref{fig:geometria}. One
consists of a $3\times3$ grid of pillars, and the other of a single pillar,
denoted as $1 \times 1$ in the following. Periodic boundary conditions are
applied in the $x$ and $y$ directions.  The pillars are $\approx 18~\sigma$ tall and
$8~\sigma$ thick, and are at a distance of $\approx 11~\sigma$ from each other.

\subsection{Collective variables and the Minimum Free Energy Path}
Physical processes and chemical reactions in complex environments are
conveniently described in terms of collective variables, i.e., a set of
$n_{CV}$ observables $\bm \phi \equiv \{\phi_1(\bm r),...,\phi_{n_{CV}}(\bm
r)\}$ that are function of the positions $\bm r = \{r_1,...,r_{n_f}\}$ of the
fluid particles.  From a macroscopic perspective, the Cassie-Wenzel transition
can be described by the advancement of the liquid/vapor meniscus $\Sigma_{lv}$
(sharp interface models) and its twisting and bending.  \cite{lv2014b,butt2013}
An alternative macroscopic approach, the (classical) density functional theory
(DFT), \cite{evans1992,lowen2002} considers the fluid density field $\rho(\bm
x)$ as the fundamental quantity to describe
wetting.~\cite{ren2014,pashos2015,panter2016} It is therefore rather natural to
use the coarse-grained density field as CV in atomistic rare event
calculations. In practice, the region of the simulation box comprising the
pillars is divided into $n_{CV}$ parallelepipedic cells, and for each cell one
counts the number of fluid particles inside it: 

\begin{equation}
\phi_k(\bm r) = {1 \over \Delta V} \sum_{i=1}^{n_f} \chi_k (\bm
r_i)\mbox{, }\quad k=1,...,n_{CV} \,\, \mbox{,} 
\label{eq:CV}
\end{equation}

\noindent where $i$ runs over the $n_f$ fluid particles,
$\chi_k$~\footnote{$\chi_k(\bm r)$, the characteristic function, is
equal to $1$ if $\bm r$ is in $k$-th cell and $0$ otherwise. $\chi_k(\bm r)$
can be expressed as the product of two Heaviside step functions per Cartesian
direction: $\chi_k(\bm r) = \prod_{l=x, y, z} \Theta(r_l - R_l^{k,b})\left
(1-\Theta(r_l - R_l^{k,e}) \right )$, with $R_l^{k,b}$ and $R_l^{k,e}$ the
\emph{b}egin and \emph{e}nd of the k-th cell in the $l$ direction, and  $r_l$
the component of the particle's position in the same direction. In practice,
the Heaviside step function is replaced by a smooth approximation, in this
case, a Fermi function. We have found this choice, which is consistent with the
literature,~\cite{miller2007} convenient, but others are possible.} is the
(smoothed) characteristic function of the $k$-th cell, and $\Delta V$ is the
volume of the cell. 

In a quasi-static representation of the wetting process a key quantity is the
probability density $p_{\bm \phi}(\bm N)$ that the collective variables $\bm
\phi$ take on the values $\bm N \equiv \{N_1,...,N_{n_{CV}}\}$.  $\bm N$
corresponds to a macroscopic state of the system, i.e., the Cassie, Wenzel or
any intermediate state visited during the transition. $p_{\bm \phi}(\bm N)$ is
frequently expressed in the logarithmic form and in thermal energy units, the
Landau free-energy  $\Omega(\bm N)$:

\begin{eqnarray}
\Omega(\bm N) &=& \beta^{-1} \log p_{\bm \phi}(\bm N) 
\label{eq:Landau} \\
&=& \beta^{-1} \log
\int d\bm r\, m(\bm r) \prod_{k=1}^{n_{CV}} \delta(\phi_k(\bm r)-N_k) \nonumber
\end{eqnarray}

\noindent where $\beta= 1/(k_B T)$ is the inverse thermal energy and
$\delta(\cdot)$ is the Dirac delta function. The probability density function
$p_{\bm \phi}(\bm N)$ is expressed in terms of the microscopic measure $m(\bm
r)$. High-probability states, i.e. Cassie and Wenzel, are global or local
minima of $\Omega(\bm N)$. 

\begin{figure}%[h!]
        \centering
        \includegraphics[width=0.45\textwidth]{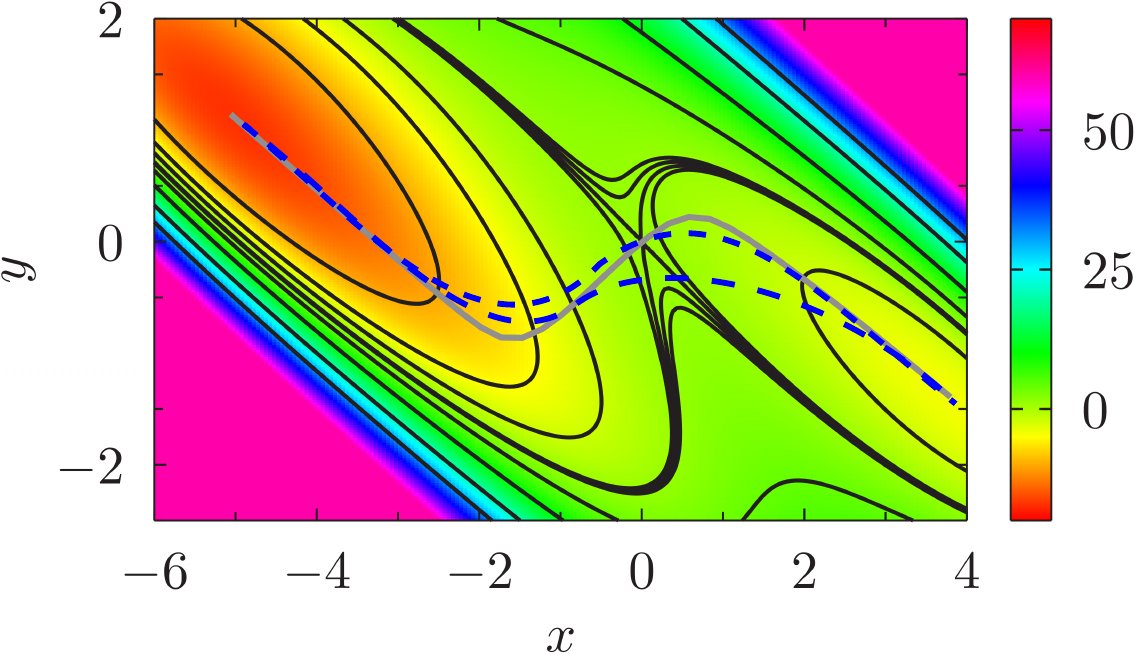}
				\caption{Multiple transition paths connecting the stable and
				metastable states (absolute and local minima) in an illustrative 2D free
			energy landscape. The grey line represents the most probable path,
		the outcome of a string calculation, and the blue dashed lines
	represent two alternative paths of lower probability.}
        \label{fig:Cartoon2D}
\end{figure}

$p_{\bm \phi}(\bm N)$ allows to gather information on the mechanism and
kinetics of the process. \footnote{Indeed, to obtain a complete understanding
of the transition mechanism and kinetics one needs also the committor function
$q_{\bm \phi}(\bm N)$, i.e. the probability that a trajectory in $\bm \phi(\bm
r) = \bm N$ reaches the final state first.  In the string method discussed in
this section, which is based on the backward Kolmogorov equation, the committor
function is taken into account implicitly.}$^,$\cite{vanden2006transition} In
principle, one could run a long MD and compute $p_{\bm \phi}(\bm N)$ from the
frequency of the associated events. However, this is not possible when the
stable and metastable states of the system are separated by free-energy
barriers considerably higher than the thermal energy. Indeed, the time required
for the system to visit the relevant regions of the CV space, including the
low-probability region corresponding to the transition state, is too large for
any present and foreseeable supercomputer. In these cases, one usually resorts
to advanced simulation techniques that enhance the sampling of regions of low
probability which are relevant for the transition. Here we use the string
method in CVs.~\cite{maragliano2006} At variance with other methods, the string
method does not require to reconstruct the free-energy landscape in the entire
CV space. Rather, the objective of the string method is to identify the most
probable path for the activated process -- the wetting path in the present case
which is related to a sequence of \emph{snapshots} of density fields of the
fluid along the Cassie-Wenzel transition. The fluid can follow many wetting
paths among which the one identified by the string method is the most probable
one. In the presence of large barriers other paths significantly different from
the one identified by the string method have a negligible probability to be
followed by the system. An illustration of multiple possible transition paths
for a simple 2D free energy landscape is shown in Fig.~\ref{fig:Cartoon2D}.
The advantage of the string method as compared to sampling techniques aimed at
reconstructing the entire free energy landscape is that its computational cost
scales only linearly with the number of CVs, while most of the other methods
have an exponential dependence on the number of CVs. \cite{laio2005} The string
method is, therefore, well suited to deal with the large set of CVs employed in
this work.

The most probable path $\left \{\bm N(\alpha) \right \}$, with the path
parametrization $\alpha\in [0,1]$, satisfies the condition
  
\begin{equation}
\left [ \hat M(\bm N(\alpha)) \nabla_N \Omega(\bm N(\alpha)) \right ]_\perp = 0 
\label{eq:MFEP}
\end{equation}

\noindent where $\perp$ indicates that Eq.~\ref{eq:MFEP} refers to the
component of $\hat M(\bm N(\alpha)) \nabla_N \Omega(\bm N(\alpha))$ orthogonal
to the path.  $ \nabla_N \Omega(\bm N)$ is the gradient of the free energy and
$\hat M(\bm N)$ is the metric matrix of elements 
\begin{equation}
\hat M_{ij}(\bm N) = \int d\bm r \nabla_r \phi_i(\bm r) \cdot \nabla_r
\phi_j(\bm r)\,\, m(\bm r) \prod_{k=1}^{n_{CV}} \delta(\phi_k(\bm r)-N_k).
\end{equation} 

\noindent$\hat M(\bm N)$ is associated to the change of variables from $\bm r$
to $\bm \phi$. \cite{maragliano2006} $\alpha$ is the independent variable of
the parametric representation of the wetting path $\bm N(\alpha)$. Here we
adopt the normalized arc length parametrization of the curve, i.e.,  $\alpha$ is
the fractional length of the arc of the curve between the initial and present
points.  Thus, in terms of the parameter $\alpha$, $\bm N(0)$ and  $\bm N(1)$
are the coarse grained densities at the Cassie and Wenzel states, respectively.
The physical meaning of Eq.~\ref{eq:MFEP} is that along the most probable
path $\bm N(\alpha)$ the \emph{effective force} $-\hat M(\bm N) \nabla_N
\Omega(\bm N)$ has zero components orthogonal to the path.

In practice, the path $\bm N(\alpha)$ is discretized in $L = 64$ images, $\bm
N(\alpha_1),\dots,\bm N(\alpha_L)$, with a constant distance between
neighboring images, 
\begin{eqnarray}
\label{eq:arclengthParam}
\delta_{\bm N} &=&   {\sqrt{ \sum_{k=1}^{n_{CV}}  (N_k(\alpha_i) -
N_k(\alpha_{i+1}))^2} \over \sum_{i=1}^{L-1} \sqrt{ \sum_{k=1}^{n_{CV}}  (N_k(\alpha_i) - N_k(\alpha_{i+1}))^2}}  \nonumber  \\
  &=& 1/(L-1) \ .
\end{eqnarray}
  
\noindent The most probable path is obtained using an iterative algorithm, the
string method~\cite{weinan2002} (see App.~\ref{app:string}), which, starting
from a first guess, produces a path satisfying the discretized version of
Eq.~\ref{eq:MFEP}. The main ingredients of Eq.~\ref{eq:MFEP}, $\nabla_N
\Omega(\bm N)$ and $\hat M(\bm N)$, can be computed using, for example, RMD
(Sec~\ref{sec:RMD}). 

The first guess path is obtained from a high pressure $P = 0.6 ~
\epsilon\sigma^{-3}$ wetting trajectory.  The wetting barrier decreases with
the pressure and, at this pressure, a wetting event has been observed over a
timescale of $\sim 10^5$ timesteps.  Along this spontaneous wetting trajectory
one extracts a number of configurations ($64$ in the present case) and computes
the associated coarse-grained density field, which represent the first-guess
path.  This approach, in which the string method is initialized from a
continuous atomistic trajectory, prevents potential problems in the calculation
of the distance $\delta_{\bm N}$ between neighboring images due to the
translational symmetry of the system: for example, in the simple definition of
the distance used here, two configurations corresponding to a translation of
the density field by one or more period in the $x$ and/or $y$ direction would
result into two different values of $\delta_{\bm N}$, even though they
represent the same physical system.  The correct choice between these two
configurations is the one which is the push-forward in time of the previous
one. Since, however, the string is a local optimization algorithm
(App.~\ref{app:string}), the initialization procedure \emph{via} a continuous
dynamics described above leads automatically to the selection of the proper
density configuration $\bf N$ among its symmetric equivalents.

Once the iterative string algorithm is converged and the most likely path is
known, one can obtain the associated energetics  by computing the line integral
of $\nabla_N \Omega(\bm N)$ along the path, thus obtaining $\Omega(\alpha)$.
Usually the free energy along the wetting is reported as a function of the
liquid fraction $\Phi(\alpha) = (n(\alpha) - n_C)/(n_W - n_C)$, with
$n(\alpha)$, $n_C$, and $n_W$ the total number of particles in the textures
(the region used to define the CVs of Fig.~\ref{fig:geometria}) at the present,
Cassie and Wenzel states, respectively. Thanks to the monotonic relation
between the fractional arc-length and liquid fraction (see Fig. S4 in the
Supporting Information), it is possible to report the free energy in the more
convenient and illustrative parametrization $\Phi$.\footnote{We
remark that the passage from the coarse-grained density field $\bm \phi$
to the liquid volume fraction in the corrugation $\Phi$ is a
non-invertible map, i.e., one has one and only one value of $\Phi$
for each value of the density field $\bm \phi$ but the opposite --that there is
only one value of the density field $\bm \phi$ corresponding to the
liquid volume fraction $\Phi$-- is false. In other words, the fact that
there is a linear relation between the fractional arc-length, and then
the associated density field $\bm \phi$, and $\Phi$ does not mean that
$\Phi$ and $\bm \phi$, i.e., CVs at different level of coarse graining,
are equivalent for describing the wetting process. Indeed,
the observation that $\Phi$ grows monotonically with the fractional
arc-length just means that the wetting process occurs with a continuous
increase of volume of the liquid in the cavity, a fact that is not
obvious but intuitive.}

When one uses a single CV, i.e., when the coarse-grained density is
replaced by the average density in the relevant domain (the yellow box
of Fig.~\ref{fig:geometria}), the path is trivial and consists in the
increase (or decrease) of the value of the single CV. With a single CV
the free-energy landscape in which the most probable path is sought for
is 1D; in this case the string method boils down to the simpler RMD.

Generally speaking, the choice of CVs (the level of coarse graining of the
density field in the present case) can affect the qualitative and quantitative
representation of the transition path.  Well established tests exist to
validate the quality of the CV set at hand, e.g., the committor
test.~\cite{bolhuis2002,maragliano2006} This test consists in numerically
computing the probability that trajectories starting from the transition
interface reach the products before  the reactants and check that the
distribution is  peaked around $50~\%$.  The transition interface is locally
defined as a plane normal to the path passing through the maximum of the free
energy (transition state).  However, performing the committor test is very
expensive. In the case of wetting, due to the slow evolution of the fluid
toward the Wenzel or Cassie state from the putative transition state, the
committor test has been rarely performed.~\cite{giacomello2015} Here we
introduce a simpler test allowing to check whether the CV set satisfies minimal
necessary conditions, namely that the path does not present unphysical
discontinuities, i.e., jumps of the density during the activated process, which
obviously cannot be present in a process resulting from continuum and atomistic
dynamics. 

\subsection{Restrained MD and related simulation techniques}
\label{sec:RMD}
In order to perform one iteration of the string algorithm one needs to compute
$\nabla_N \Omega(\bm N)$ and $\hat M(\bm N)$ at the set of points $\{\bm
N(\alpha_i)\}_{i=1,\ldots,L}$.  In this section we show how these quantities
can be computed by RMD.
 
Consider Eq.~\ref{eq:Landau} defining the Landau free energy and replace the
Dirac delta functions on the r.h.s. by smooth approximations of Gaussian form
$g_\lambda(\phi_k(\bm r)-N_k) = \sqrt{2\pi/(\beta \lambda)} \exp[-\beta
\lambda/2\,(\phi_k(\bm r)-N_k)^2]$. Within this approximation each component of
the gradient of the free energy reads

\begin{eqnarray}
{\partial \Omega(\bm N) \over \partial N_i} &\approx& {\partial \Omega_\lambda(\bm N) \over \partial N_i} = \\ \nonumber
&=& {\int d\bm r \,\, \lambda\, (\phi_i(\bm r)-N_i)\, m(\bm r)
	\prod_{k=1}^{n_{CV}} g_\lambda(\phi_k(\bm r)-N_k) \over
\int d\bm r \,\, m(\bm r) \prod_{k=1}^{n_{CV}} g_\lambda(\phi_k(\bm r)-N_k)} \\ \nonumber
&=& {\int d\bm r \,\, \lambda (\phi_i(\bm r)-N_i)\, p(\bm r | \bm N)} \ .
\label{eq:gradLandau}
\end{eqnarray}

\noindent Thus, ${\partial \Omega(\bm N) / \partial N_i}$ is approximated by
the expectation value of the observable $\lambda (\phi_i(\bm r)-N_i)$ over the
conditional probability density function $p(\bm r| \bm N)$.  Here, we have
set $\lambda = 0.2$, which is a trade-off between the convergence of $
{\partial \Omega_\lambda(\bm N) / \partial N_i} $ with $\lambda$ and the
statistical error of the mean force (see Supporting Information Fig. S2(a)).
The suitability of this value of $\lambda$ has also been tested for convergence
by analizing the behaviour of $\Omega$ with $\lambda$ (Fig. S2(b)). 

If $m(\bm r)$ is the measure of the canonical ensemble, $m(\bm r)
\prod_{k=1}^{n_{CV}} g_\lambda(\phi_k(\bm r)-N_k)$ can be cast in the form
$\exp\left(-\beta (V(\bm r) + \sum_{k=1}^{n_{CV}} \lambda/2 (\phi_k(\bm
r)-N_k)^2)\right)$, which suggests that the conditional probability density can
be sampled by a constant temperature MD driven by the \emph{augmented}
potential $\tilde V(\bm r; \bm N) = V(\bm r) + \sum_{k=1}^{n_{CV}} \lambda/2
(\phi_k(\bm r)-N_k)^2$, the so-called \emph{Restrained} MD. Indeed, this
argument can be extended to the isothermal-isobaric ensemble, provided that one
uses a molecular dynamics suitable to sample this ensemble. 

In practice, ${\partial \Omega(\bm N) / \partial N_i}$ is computed as the time
average of $\lambda (\phi_i(\bm r)-N_i)$ along the RMD.  $\hat M(\bm N)$ can
also be computed as a time average along the same RMD, this time considering
the observable $\nabla_{\bm r} \phi_i(\bm r) \cdot \nabla_{\bm r} \phi_j(\bm
r)$.

We remark that for a single CV one does not need the string method to
compute the wetting/dewetting path, which is just a segment containing
the values of the CV describing the process from the initial to
the final state. Thus, in this case RMD allows to compute directly the
free energy via the numerical integration of  ${d \Omega(N) / d N}$. In
this sense, sometimes we refer to RMD as a proxy for a single-CV string.

\section{Results and discussion}
\label{sec:results}
This section is divided in three subsections, i) \emph{$3\times 3$ pillars
system} (Sec.~\ref{sec:grande}) and ii) \emph{$1\times 1$ pillar system}
(Sec.~\ref{sec:piccolo}), in which we report and discuss results of the effect
of different levels of coarse graining of the density field on the wetting path
and its energetics for each sample independently, and iii) \emph{Size effects:
comparison between $3 \times 3$ and $1 \times 1$ pillars systems}
(Sec.~\ref{sec:confronto}), in which we discuss the effect of the size of the
surface sample on the results.

%%%%%%%%%%%%%%%%%%%%%%%%%%%%%%%%%%%%%%%%%%%%%%%%%%%%%%%%%%%%%%%%%%%%%%%%%%%

\subsection{$\bm 3 \bm \times \bm 3$ pillars system}
\label{sec:grande}

\begin{figure*}%[h!]
        \centering
        \includegraphics[width=1.0\textwidth]{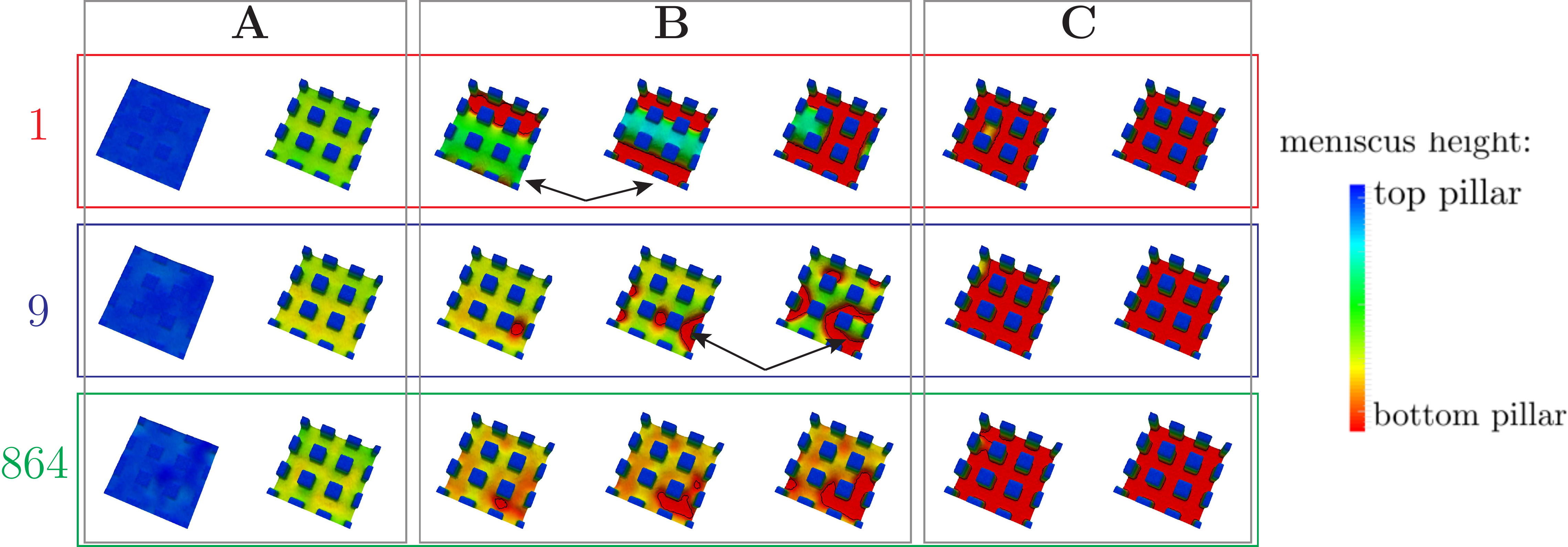}
				\caption{Configurations of the  meniscus along
the wetting process for $1$ (red frame), $9$ (blue frame), and $864$ (green
frame) CVs. The color code (bottom left) helps identifying the distance of the
meniscus from the bottom wall: blue when the meniscus is at the top of the
pillars, red when it touches the bottom, and green in between.  In the figure
we also report the triple line formed by the liquid/vapor interface with the
bottom wall (black).(see Supporting Information for Publication for the
movies of the wetting trajectories, files 3x3pillarCV864.gif, 3x3pillarCV9.gif,
3x3pillarCV1.gif).}
        \label{fig:percorso}
\end{figure*}

Figure~\ref{fig:percorso} shows the evolution of the liquid meniscus along the
collapse of the Cassie state for $n_{CV}=1$ ($1 \times 1 \times 1$ mesh
with a cell of $\approx 56 \times 56 \times 29\,\, \sigma$), $n_{CV}=9$
($3 \times 3 \times 1$ mesh with cells of $\approx 18.7 \times 18.7 \times
29\,\,\ \sigma$), and $n_{CV}=864$ ($12 \times 12 \times 9$ mesh with
cells of $\approx 4.6 \times 4.6 \times 4.8\,\, \sigma$
- Fig.~\ref{fig:geometria}). A color coding is applied to help visualizing the
  distance of the meniscus from the bottom wall of the surface.

The meniscus is identified using the Gibbs dividing surface, i.e., the locus of
points where the fluid density coincides with the mean value between the liquid
$\rho_l$ and vapor $\rho_v$ bulk values, $(\rho_l-\rho_v)/2 \approx 0.375~\sigma^{-3}$.
Thus, identifying the meniscus from an ensemble of atomistic configurations
requires computing a coarse-grained density field, $\rho(\bm x)$. For this
computationally inexpensive post-processing operation we use a finer level of
coarse graining than that used for the CVs, namely a $56 \times 56 \times 29$
points grid (cell $\approx 1\times 1 \times 1~\sigma$).

\begin{figure}%[h!]
        \centering
        \includegraphics[width=0.45\textwidth]{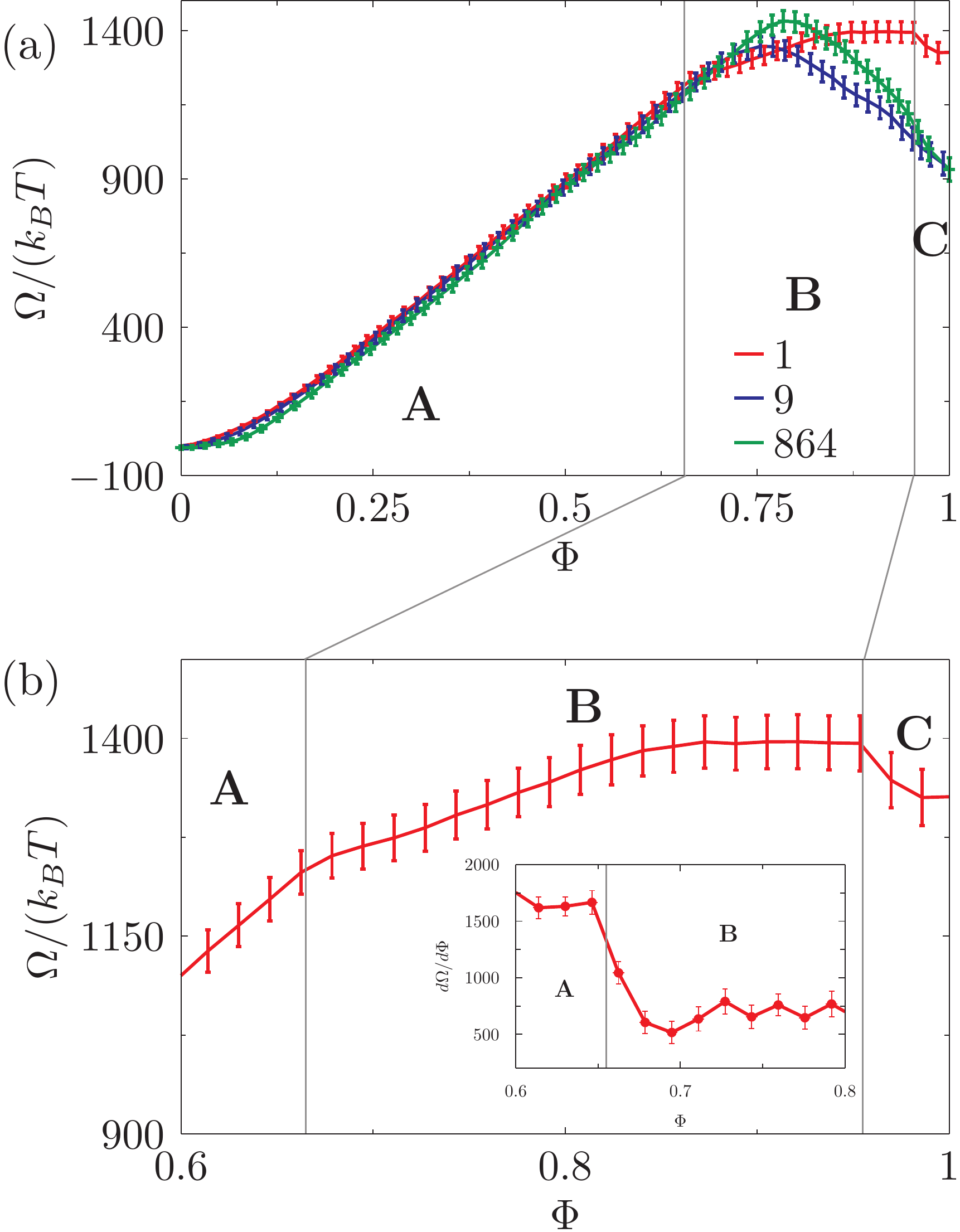}
				\caption{a) Free-energy profiles for  $1$, $9$,
and $864$ CVs. Error bars are computed via the procedure described in the
Supporting Information. b) Zoom of the free-energy profile for $1$
CV in region B and C highlighting the change of slope of this curve associated
to the morphological transitions discussed in the text.  The mean force
relative to the first change of slope at the transition between the A and B is
reported in the inset. Inspection of the mean force for $1$ CV shows a double
discontinuity in this domain. In the Supporting Information the 1 CV
free-energy profile and mean force are reported with a finer discretization of
the transition path in the B and C domains, confirming the results discussed in
the main text.
\label{fig:profili}}
\end{figure}

The analysis of the meniscus shapes along the wetting process reveals that the
collapse mechanism consists of three steps (A-C) (Fig.~\ref{fig:percorso}). In
step A, the liquid, initially pinned at the top of the pillars, starts to
progressively fill the inter-pillar space, with the meniscus assuming an
essentially flat conformation parallel to the bottom wall. This step is similar
for all the CVs sets, and corresponds to the linear increase of the free energy
in Fig.~\ref{fig:profili}(a). 

In step B, which includes the transition state, the liquid touches and spreads
over the bottom wall.  The existence of a maximum of the free energy in this
domain can be explained using an intuitive argument: i) while the liquid slides
down along the pillars with a flat meniscus the liquid-solid surface increases
resulting in an increase of the energetically unfavorable solid-liquid 
contribution; ii) when the liquid  starts spreading over the bottom
wall the two liquid-vapor and vapor-solid interfaces are replaced by a single
liquid-solid one, which results in a reduction of the free energy. Thus, the
free energy switches from an increasing to a decreasing trend when 
the meniscus comes in contact with the bottom wall.

Step B of the wetting path is significantly different in the three cases
(Fig.~\ref{fig:percorso}). With $1$ CV, the liquid touches
the bottom wall by completely filling the space between two rows of pillars.
Due to the 2D periodicity, this amounts to having an infinite strip of liquid
touching the bottom of the textured surface parallelled by an infinite strip of
vapor, with the corresponding meniscus facing the bottom wall at a distance of
ca. $6~\sigma$. The process proceeds with the liquid  progressively invading
neighboring rows or squares  of pillars.

For both $9$ and $864$ CVs, step B of the wetting process starts in a similar
way, with the liquid touching the bottom  between two pillars at a single
point.  However, wetting evolves in a rather different way. In the case of
$864$ CVs the liquid \emph{percolates} in neighboring inter-pillar regions;
with $9$ CVs it wets disjoint regions.

A remarkable aspect of step B with $1$ and $9$ CVs is
that both present \emph{discontinuities} in the path corresponding to
abrupt changes of the meniscus configuration (Fig.~\ref{fig:percorso}). 
For example, in the case of $1$ CV the
filling of the corrugations proceeds with the liquid suddenly wetting
entire squares or rows among pillars. Concerning the $9$ CVs
case, comparing the second and third snapshots of step B of the
wetting path one notices that the bottom wall between the right-low
pair of pillars is initially wet by the liquid, while it gets dry in the
next image (see arrows in Fig.~\ref{fig:percorso}).  As we discuss more
extensively below, the origin of these unphysical discontinuities in the
paths is the inability of $1$ and $9$ CVs to distinguish among different
meniscus configurations: more than one configuration of the meniscus is
possible for a given value of the CVs. 

Step C of the path is similar for all the cases: a rarefied liquid
region between pairs of pillars transforms into bulk liquid, which
eventually completely wets the bottom wall. Videos of the wetting
process for the three cases are available in the Supporting Information.

Concerning step B of the wetting path, which includes the kinetic bottleneck of
the process -- the transition state, an important question is whether the paths
identified with the three sets of CVs are all physically sound. Before
addressing this question let us revise the concept of transition path in rare
event simulations. The path is described as a parametric (discretized) curve in
the space of the CVs.  The parametric representation is, indeed, only a
convenient alternative to the representation of the path as a function of
time.~\cite{maragliano2006} Thus, for systems evolving according to continuous
dynamics, one would expect that observables change smoothly between successive
images of the discretized path. Of course, in all collective-variable-based
rare event techniques, the CVs used to follow or accelerate the process change
by a prescribed value between images; for example, in the present string
calculations the \emph{distance} in the value of CVs between successive images
is constant by construction, Eq.~\ref{eq:arclengthParam}. However, one
expects that this property holds for all the fundamental variables of the
system. In particular, in the case of wetting of textured surfaces, which is
well described by the (classical) density functional
theory,~\cite{ren2014,tretyakov2016} a necessary condition for the CVs to be
suitable to describe the process is that the density field (computed,
virtually, on an infinitely resolved grid) changes in a continuous way along
the path regardless of the degree of coarse graining considered. To verify
whether this condition is met, we computed the nondimensional Euclidean
distance between (ensemble averaged) density fields $\rho(\bm x, \Phi_i)$
computed on a very fine $56 \times 56 \times 29$ mesh (cell $ \approx
1 \times 1 \times 1~\sigma$) at successive images $i$ along the paths
identified via the three sets of CVs:
\begin{eqnarray}
  \label{eq:deltarho}
  \delta_{\rho} (\Phi_i) &=& \frac{\sqrt{ \sum_{l,m, n}  (\rho(\bm x_{l,m,n}, \Phi_i) - \rho(\bm x_{l,m,n}, \Phi_{i+1}))^2}}{\rho_l} 
	\mbox{ ,}
\end{eqnarray}
where the normalization is performed with respect to the bulk liquid density
$\rho_l$, and the sum $\sum_{l,m, n}$ runs over the cells indices of the very
fine mesh, which is the same for all the CVs. We remark that, apart for
the normalization and the mesh used in Eq.~\ref{eq:deltarho}, $\delta_\rho$
is equivalent to the normalized arc length $\delta_N$
(Eq.~\ref{eq:arclengthParam}). The value of $\delta_{\rho} (\Phi_i)$
far from discontinuities depends on the absolute arc length of the path in the
CV space and on the number of images in which it has been discretized. Thus,
there is no trivial relation between the number of CVs and the value of
$\delta_{\rho} (\Phi_i)$. In practice, for the case of the wetting of pillared
surfaces, far from discontinuities $\delta_{\rho} (\Phi_i)$ takes similar
values for all the considered sets. 

\begin{figure}%[h!]
        \centering
        \includegraphics[width=0.45\textwidth]{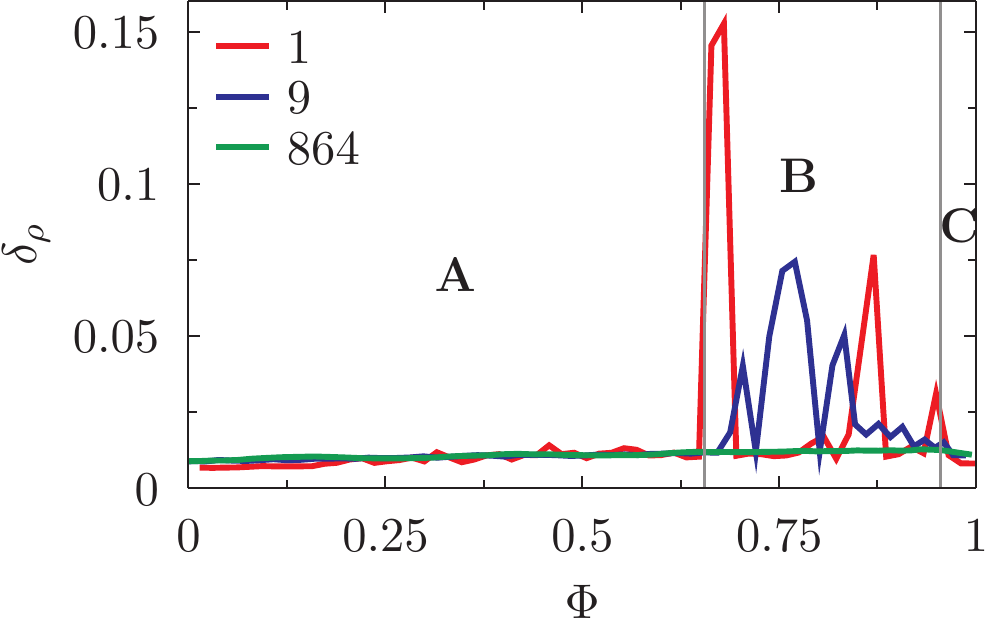}
                                \caption{
Nondimensional Euclidean distance $\delta_{\rho} (\Phi)$ between density
fields at successive points along the string $\rho(\bm x, \Phi)$ (see
Eq.~\ref{eq:deltarho}). Consistently with Fig.~\ref{fig:percorso},  red, blue
and green curves refer to  $1$,  $9$, and $864$ CVs, respectively.}
        \label{fig:deltarho}
\end{figure}

It is observed (Fig.~\ref{fig:deltarho}) that in steps A and C, $\delta_{\rho}
(\Phi_i) \approx 0.01$ independently of the number of CVs, indicating that
there are no major differences in the continuity of the paths identified by the
three CVs in these regimes. On the contrary, in step B the paths obtained with
$1$ and $9$ CVs show discontinuities with a sizeable increase of $\delta_{\rho}
(\Phi_i)$, which reaches $\approx 0.15$ and $\approx 0.07$ in the first and
second case, respectively. These results suggest that $1$ and $9$ CVs are
insufficient to describe the wetting process because, in the most important
region of the path --the one determining the kinetics of the process-- they are
inadequate for describing the continuous trajectory followed by the liquid
wetting the surface textures. 

The difference in the wetting paths is reflected in the difference among the
free energy profiles (Fig.~\ref{fig:profili}(a)).  In the case of $1$ CV we
observe an extended, relatively flat domain of the free energy profile in step
B. Indeed, careful analysis shows that this region is composed of two linear
segments with a different slope (Fig.~\ref{fig:profili}(a)). The discontinuity of the first derivative is
observed in correspondence of the transitions from one morphology to another
(see inset of Fig.~\ref{fig:profili}(b)), namely i) from the \emph{flat}
meniscus to the liquid partly wetting the bottom wall (transition from step A
to B); ii) from the liquid completely wetting one row of pillars to two rows;
iii) the passage from the liquid wetting only part of the bottom wall, with a
layer of \emph{bulk vapor} separating the liquid meniscus from the bottom, to
the liquid wetting the bottom wall, still containing small vapor bubbles 
between pairs of pillars (transition from step B to C). These three points
coincide with the three sharp peaks of $\delta_\rho$.  The profile is more
regular in the case of $9$ and $864$ CVs, with a well defined maximum of the
free energy between the Cassie and Wenzel states. 

The quantitative comparison of the three free-energy profiles reveals two
important aspects. First, the difference of free energy between the Cassie and
Wenzel states, $\Delta \Omega_{CW}$, which determines their relative stability,
depends on the CVs. $9$ and $864$ CVs yield consistently $\Delta \Omega_{CW}=
940 \;k_BT$, while the $1$ CV case is ca.~$400~k_BT$ higher. This discrepancy,
in turn, affects the barrier of the Wenzel-Cassie transition, reducing its
estimate by a corresponding amount in the case of $1$ CV. Second, also the
Cassie-Wenzel free-energy barrier $\Delta \Omega_{CW}^\dagger$ changes from one
CVs set to another. In particular, $\Delta \Omega_{CW}^\dagger$ is
approximatively $40$ and  $80~k_BT$ lower for the $1$ and $9$ CVs,
respectively, as compared to the $864$ CVs case.

We remark that the differences in the free energy profiles with the three
CV sets is not due to an insufficient discretization of the RMD and string
paths. To show this, we repeated the RMD calculation in B and C  with three
times the number of points (see Supporting Information Fig. S5(b)). The
original and more accurate profiles show minor quantitative differences. To
further confirm that the difference in the free energy profiles does not depend
on an insufficient discretization of the path but on the difference of the CV
sets employed, in the region across one of the changes of slope of $\Omega$ for
1 CV we increased the number of point by one order of magnitude (Fig. S5(a)).
Also in this case we observed minor quantitative differences in both the free
energy profile and mean force.

\begin{figure*}%[h!]
\centering
\includegraphics[width=1.0\textwidth]{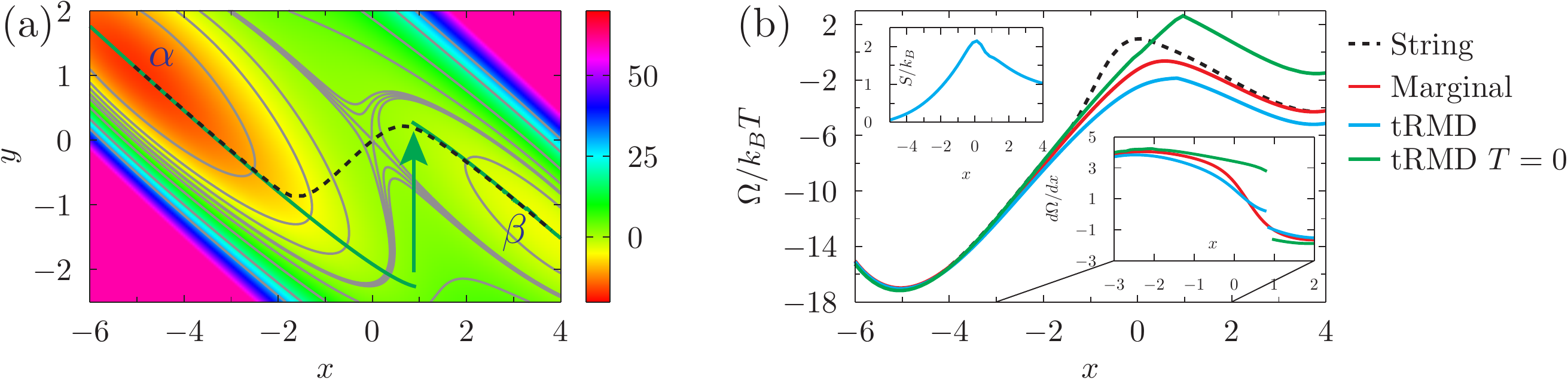}
\caption{a) 2D Polynomial potential $v(x,y)/(k_BT) =  ((0.6 x+0.8
	y)^2-1)^2 - 2.0 (0.8 x-0.6 y) (0.6 x+0.8 y) + 0.01 (0.8 x-0.6 y)^4 +
	1.5 (0.6 x+0.8 y)^3$ in the two arbitrary variables $x$ and $y$. 
The potential is characterized by two partly overlapping attractive basins,
i.e., depending on the position in $y$ the force along $x$ can be attractive
toward either minimum. The color map helps identifying minima and saddle points
of the potential. The dashed line represents the string path with two
variables; the green line represents the two
branches of the discontinuous zero temperature tRMD path computed by applying a
restraint on the $x$ variable and starting from basin $\alpha$. The arrow indicates the point in which the
system jumps from one branch to the other.  b) Comparison among energy profiles
obtained with different numbers of variables, methods and temperatures. The
dashed line is the energy profile along the two-variables string reported as a
function of $x$. The red curve is the \emph{exact} projection of the 2D
potential on the variable $x$, $v(x) = -k_BT \log \int dy \exp[-\beta v(x,y)]/
\int dy dx \exp[-\beta v(x,y)]$ and $\beta = 1$. The green
curve is the free energy obtained by integrating the mean force calculated from
a zero temperature tRMD simulation, and the cyan curve is the
finite temperature equivalent at $k_B T=1$. 
In the inset at the bottom-right of panel b is
reported the mean force $d\bar v(x)/dx$ for the \emph{exact} (red), zero
(green) and finite temperature (cyan) tRMD
cases. This inset illustrates that in both tRMD cases the forces at the
transition between the two branches of the path are discontinuous, with the
discontinuity reducing at finite temperature.  In the inset at the top-left of
panel b  the entropy is reported along the finite temperature tRMD path.}
\label{fig:toy}
\end{figure*}

\subsubsection{Dependence of $\Delta \Omega_{CW}$ on the CVs}
The theoretical arguments developed in App.~\ref{app:1vs2CVs} indicate that,
the values of $\Delta \Omega_{CW}$ should be close for the three CVs; the
present results conflict with this theoretical prediction.  The origin of this
mismatch can be illustrated considering a model 2D polynomial
potential $v(x,y)$ which has an analytical expression and on which different
CVs can be analyzed in detail (Fig.~\ref{fig:toy}(a)).  In panel b) the profile
of the potential along the 2D string path (dashed black line of
Fig.~\ref{fig:toy}(a)) is reported; this is the most probable path joining
states $\alpha$ and $\beta$.  We also consider the 1D counterpart of the
potential, which is, in our analogy, the equivalent of the free-energy profile
obtained with higher coarse graining; this 1D potential is obtained by
projecting $v(x,y)$ along the $x$ axis according to Eq.~\ref{eq:Landau}:
$v(x) = -k_BT \log p(x)$, where $p(x) =  \int dy \exp[-\beta v(x,y)]/ \int
dy dx \exp[-\beta v(x,y)]$ is the marginal probability density
function.\footnote{The marginal $p(x)$ is the probability density function
that $x$ takes a given value while $y$ takes any value. Of course the concept
of marginal is not limited to the case of two random variables but can be
extended to the case of any number of random variables, and refers to the event
that a subset takes a given value while the rest of the variables take any
value.}  Figure~\ref{fig:toy}(b) shows that $\Delta v_{\alpha\beta}$, the
energy difference between the two minima, is indistinguishable if computed
along the string path or using  the 1D potential $v(x)$; the differences
\emph{along} the path will be analyzed later on when the barriers are
considered.

We further compute $\Delta v_{\alpha\beta}$ as one would obtain from an RMD
simulation (thought RMD - tRMD) performed using $x$ as CV; this is expected to
provide a numerical approximation to $v(x)$.  In tRMD one imagines to start
from the minimum $\alpha$ increasing step by step $x$.  In finite time, for
each $x$, the system explores configurations within few $k_B T$ from the local
minimum $y^\ast(x)$ at the constrained value of $x$ belonging to the basin of
$\alpha$ (green curve in Fig.~\ref{fig:toy}(a)). When the barrier along $y$
separating the $\alpha$ and $\beta$ basins is sufficiently small the system
jumps into the other basin and successively oscillates around the other
branch of the green curve which brings to the basin $\beta$. In
Fig.~\ref{fig:toy}(a) the transition from the initial to the final branch of
the path occurs when the barrier along the $y$ direction is $1~k_BT$. From the
ensemble of $\nu$ configurations available at each $x$ one estimates the mean
gradient
\begin{equation}
\label{eq:meanGradient}
{\overline  {dv(x)\over dx} } = \frac{1}{\nu} \sum_{i=1}^\nu \frac{\partial v(x,y_i)}{\partial x}
\end{equation}
from which the free-energy profile is reconstructed by integrating with respect to $x$. 

As noted, the values of $\Delta v_{\alpha\beta}$ obtained from  $v(x,y)$ and
the projected $v(x)$ coincide, while $\Delta v_{\alpha\beta}$ is not captured
correctly by tRMD.  We remark that the relative error on $\Delta
v_{\alpha\beta}$ estimated via tRMD is $\approx 20\%$, of the same order of
magnitude of the error on $\Delta \Omega_{CW}$ with 1 CV. The reason of the
error on $\Delta v_{\alpha\beta}$ is that the mean gradient used to compute the
free-energy profile in tRMD, Eq. \ref{eq:meanGradient}, differs from the
actual one near the jump between the two valleys of the potential (see the
inset of Fig.~\ref{fig:toy}(b); additional details are discussed in
App.~\ref{app:1vs2CVs}). In particular, $\overline{ d v(x)} / dx$ obtained from
tRMD is discontinuous when the trajectory passes from one basin to the other. 

This analysis of the simple 2D potential supports our claim that the difference
in the $\Delta \Omega_{CW}$ between $1$ and $9$ or $864$ CVs is due to the
simulation protocol. The transition from one basin to another in the 2D
potential is equivalent to the  meniscus jumps observed in the wetting process
with $1$ CV. Thus, we believe that the difference in $\Delta \Omega_{CW}$
between $1$ and $9$ or $864$ CVs can be ascribed to the severe discontinuities
in the gradient of the free energy in the former case, which are induced by
the extreme coarse graining of the associated density field in the case of one
single CV. These discontinuities are shown for the case of a single collective
variable in the inset of Fig.~\ref{fig:profili}(b).

Indeed, this problem, which is also at the origin of the
hysteresis observed with other rare event techniques when one uses
insufficient or inadequate CVs, could be solved combining RMD with
parallel
tempering,~\cite{1986PhRvL..57.2607S} which allows to correctly sample the
conditional probability in the space complementary to the collective variables
(see, e.g., Refs.~\citenum{Orlandini:JournalOfStatisticalPhysics:2011} and
\citenum{PhysRevB.83.235303} for the application of this combined approach
to phase transitions). This approach would, thus, bring $\Delta \Omega_{CW}$ of
$1$ CV closer to the value computed with $9$ or $864$ CVs; however, to the best
of our knowledge, due to the high computational cost this approach has been
rarely adopted in the context of wetting transition.~\cite{giacomello2012}

\subsubsection{Dependence of the wetting barrier on the CVs}
In the following, when comparing the free energy barrier of different CVs,
we will first focus on the \emph{zero temperature} limit and we will then
consider \emph{finite temperature effects}. It must be stressed that zero and
finite temperature refers to the CVs degrees of freedom, while the atoms, and
thus the free-energy landscape, are always at the physical temperature $T =
0.8~\epsilon/k_B$. $1$ CV is a subset of $9$ CVs which, in turn, is subset the
$864$ CVs. Thus, in the present context, the zero-temperature limit of the $1$
CV set means that all the remaining $863$ uncontrolled degrees of freedom to
complete the $864$-dimensional CV space take the value corresponding to the
constrained local or global minimum at the present value of the restrained CV.
An analogous argument holds for the comparison of the barriers with $1$ and $9$
CVs, and $9$ vs $864$ CVs. The reason to split the comparison in the zero
temperature limit and finite temperature effects is that this allows to carry
on an accurate theoretical analysis, bringing to some interesting conclusion on
the possibility or impossibility to establish an ordering of the barriers with
the size of the CV set.

Let us consider first the Cassie-Wenzel transition in the limit of zero
temperature, and explain the phenomenology using the model 2D polynomial
potential. Within this limit and for the $x$ variable only, the system strictly
follows the path consisting of the line of the constrained minima of the energy
laying in the $\alpha$ basin until the barrier in the orthogonal space is zero
(green line in Fig.~\ref{fig:toy}(a)); then the system follows the line of
constrained minima in the $\beta$ basin (blue line). The string in $2$
variables departs from the zero-temperature tRMD path in the intermediate
region, undergoing the transition from the initial to the final basin at a
different point. Using a different language, the trajectories with one and two
variables cross the separatrix, the surface separating the two attractive
basins, at different points. In Ref.~\citenum{maragliano2006} it is shown
that the string path crosses the separatrix at the minimum of the potential on
this surface, and this point is the actual transition state. Thus, the
zero-temperature tRMD transition state has higher or equal energy than the 2D
string transition state (see Fig.~\ref{fig:toy}(b)). We remark that this
ordering of the barriers is valid only in the zero-temperature limit, and is due to
the fact that the system keeps moving along the line of constrained minima of
the basin $\alpha$ beyond the point $x$ at which the potential of the minimum
in the basin $\beta$ would be lower. 

The argument developed above is not limited to the case of the 2D polynomial
potential but is also valid  in the case of the CVs for the wetting process
discussed in this work (see App.~\ref{app:1vs2CVs}). One notices that the
barriers with $1$ and $9$ CVs are lower than that with $864$, which conflicts
with the zero-temperature analysis above and the intuitive expectation that the
barriers should decrease as the number of CVs increases. This suggests that
there must be other effects which are not present in the zero-temperature
limit. 

At finite temperature there are two additional effects to take into account.
The first concerns the point where the transition occurs with a reduced number
of CVs. At finite temperature this is attained where  the orthogonal barrier is
of the order of several $k_BT$,~\footnote{The transition from the
initial to the final branch of the path is performed when the barrier along the
$y$ direction is zero and $1 k_BT$ for $T=0$ tRMD and tRMD, respectively. The
first condition is met at the green arrow of Fig.~\ref{fig:toy}(a). The second
condition occurs very close to this point.} i.e., when the transition time is
of the same order of magnitude of the string/RMD  simulation time. This implies
that the transition occurs before than in the zero temperature case, i.e.,
closer to the initial state, which results in a reduction of the barrier.

The second effect concerns the entropic contribution to the energetics of the
transition, which is associated to the number of states the system can explore
in the space orthogonal to the collective variables at a given point along
the wetting path. The higher is the number of collective variables the lower is
the dimensionality of the configuration space the system can explore and, thus,
the lower should be the configurational entropy at a given point along the
path.  For example, in the case of 864 CVs at each point along the string the
fluctuations allowed the density field to vary over a spatial scale smaller
than the cell in which the system is partitioned, i.e. on a spatial scale of
$\approx 4.6 \times 4.6 \times 4.8 \,\,\ \sigma$.  In the case of $9$ CVs the
density can fluctuate on a much larger spatial scale of $\approx 18.7 \times
18.7 \times 29\,\sigma$.  This effect can be better illustrated in the case of
the 2D polynomial potential. At each point along the string in $(x, y)$ the
system can take only one configuration, and the entropy is zero. On the
contrary, in the case of finite temperature tRMD the system oscillates in the
$y$ direction around the minimum $y^\ast(x)$ at the current value of $x$; the
entropy at each point along the RMD path is different from zero.

This effect is qualitatively valid in general, at all points along the
path. However, the value of the entropy depends on the profile of the potential
in the space orthogonal to the CVs at the current point along the string/RMD.
Thus, in the 2D potential the entropy of the system at a given $x$ depends on
the shape of $v(x,y)$ in the $y$ direction: if the potential is stiff the
entropy is low, if it is shallow the entropy is high. For example, the
potential in the $y$ direction is stiff at the initial and final states and
shallow at the transition state: thus the entropic contribution, $-T\Delta S$,
tends to reduce the barrier as compared to tRMD at $T = 0$ (see inset of
Fig.~\ref{fig:toy}(b)).  

The combined effect of the temperature in reducing the barrier, anticipating 
the transition and by entropic effect, is illustrated by comparing tRMD at zero
and finite temperature.~\footnote{It is worth remarking that there
is no obvious ordering among the free energy profiles of the single variable
$x$ because the value of the free energy depends on several competing factors:
the value of the free energy at the constrained minima in the two basins, the
associated curvature and the temperature (see also App.~\ref{app:1vs2CVs}).}
Indeed, Fig.~\ref{fig:toy} shows that at finite temperature the tRMD barrier
can be lower than the string one.

A similar scenario is likely to occur for the wetting process, in which $1$ and
$9$ CVs might have a higher entropy than $864$ at the transition state.
For the case of $9$ CVs this hypothesis is supported by the visual inspection
of the trajectories.  The \emph{fluctuations} of the meniscus close to the
Cassie state, in which the liquid/vapor interface is pinned at the pillars top,
is small in all cases. On the contrary, at the transition state the meniscus
shows larger fluctuations for $9$ CVs than for $864$. 

\begin{figure}%[h!]
        \centering
        \includegraphics[width=0.45\textwidth]{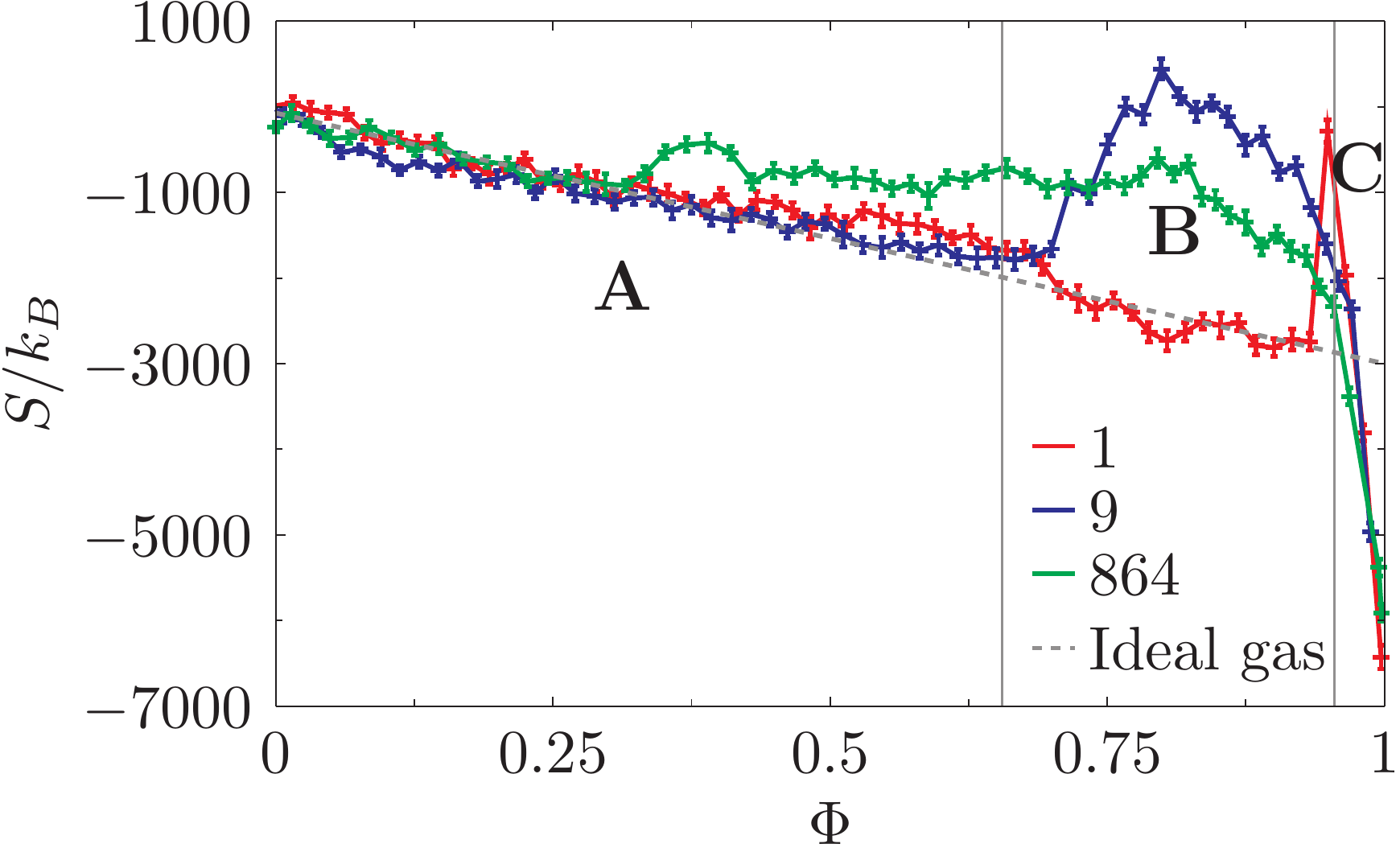}
				\caption{Entropy along the wetting path for the
same sets. The reference is chosen such that the entropy is zero in the Cassie
state. The entropy is computed from the free energy by subtracting the
enthalpy, which, in turn, is computed as the sum of the expectation value of
the Hamiltonian $\langle H \rangle$ and the $-PV$ term of the liquid and vapor
phases.  The entropy profile is noisy due to the corresponding noisy signal of
$\langle H \rangle$. The dashed line represents the entropy of the ideal gas
computed via the Sackur-Tetrode formula $S = k_B N \left ( -\log (\rho_v
\Lambda^3) + 5/2 \right )$, where $\Lambda = h / \sqrt{3\ m\ k_BT}$ is the
thermal wavelength, with $h$ the Plank constant and $\rho_v = 0.045\sigma^{-3}$
density of the gas. N is the number of particles in the vapor phase inside the
pillars which is computed as $N=V(1-\Phi)\rho_v$ where $V$ is the volume inside
the pillars.
\label{fig:entropy}}
\end{figure}

To validate this hypothesis we have computed the wetting entropy by
subtracting from the free energy the enthalpic  term, given by the average
Hamiltonian at each image of the path plus the pressure term ($S_\Phi = -1/T\
\left [ \Omega_\Phi - (\langle H \rangle_\Phi + P_l V_l + P_v V_v) \right ])$,
with $\langle H \rangle_\Phi $ denoting the ensemble average of the Hamiltonian
at the present point of  string/RMD and $P_x$ and $V_x$ the pressure and
volume, respectively, of the phase $x$, \emph{l}iquid or \emph{v}apor). It is
seen (Fig.~\ref{fig:entropy}) that the entropy of the three CVs initially
has a very similar descending trend, which is due to the reduction of the
amount of the highly entropic vapor phase  (gray dashed line in the figure). In
correspondence of the transition states both the $9$ and $864$ CV cases present
an increase of the entropy, which is, however, more pronounced for the $9$ CVs.
This, indeed, explains why the $9$ CVs set has a barrier  $80~k_BT$ lower than
the $864$ one.

The scenario is more complex for the $1$ CV case. Here, at variance with
the other two cases, in step B the entropy further decreases. We
believe that this is due to the sudden elimination of a portion of the
liquid-vapor interface, which, with its fluctuations, contributes to the
entropy of the system. This is different from the other two cases, in
which the liquid-vapor interface is eliminated gradually, and the curved
liquid-vapor interface joining the triple line at the bottom wall with
the flat part of the meniscus increases the entropy. 
At the boundary between B and C, when teh $1$ CV system changes
from a well defined liquid-vapor biphasic system to one made of a liquid
plus an extended and diffused liquid-solid interface, the entropy
suddenly increases.  The final part of the entropy profile is similar
for all cases: it decreases following the absorption of the regions of
rarefied liquid (Fig.~\ref{fig:percorso}). 

Summarizing, the level of coarse graining affects 
the wetting barrier in three different ways: i) at zero temperature a
lower degree of coarse graining increases the barrier because the
separatrix is crossed outside of its minimum; at finite temperature ii) the
crossing of the separatrix is anticipated and the barrier reduced in all
cases, but the effect is higher for more coarse grained density fields
and iii) entropic effects may further reduce the barrier for more coarse
grained density fields. However, this last argument has exceptions for
too coarse grained density fields, e.g., a single CV, which changes the
wetting mechanism toward less entropic paths. The balance between these
three effects cannot be easily predicted. Rather, they simply explain
the ordering of barriers observed with $1$, $9$ and $864$ CVs.

\begin{figure*}%[h!]
        \centering
        \includegraphics[width=1.0\textwidth]{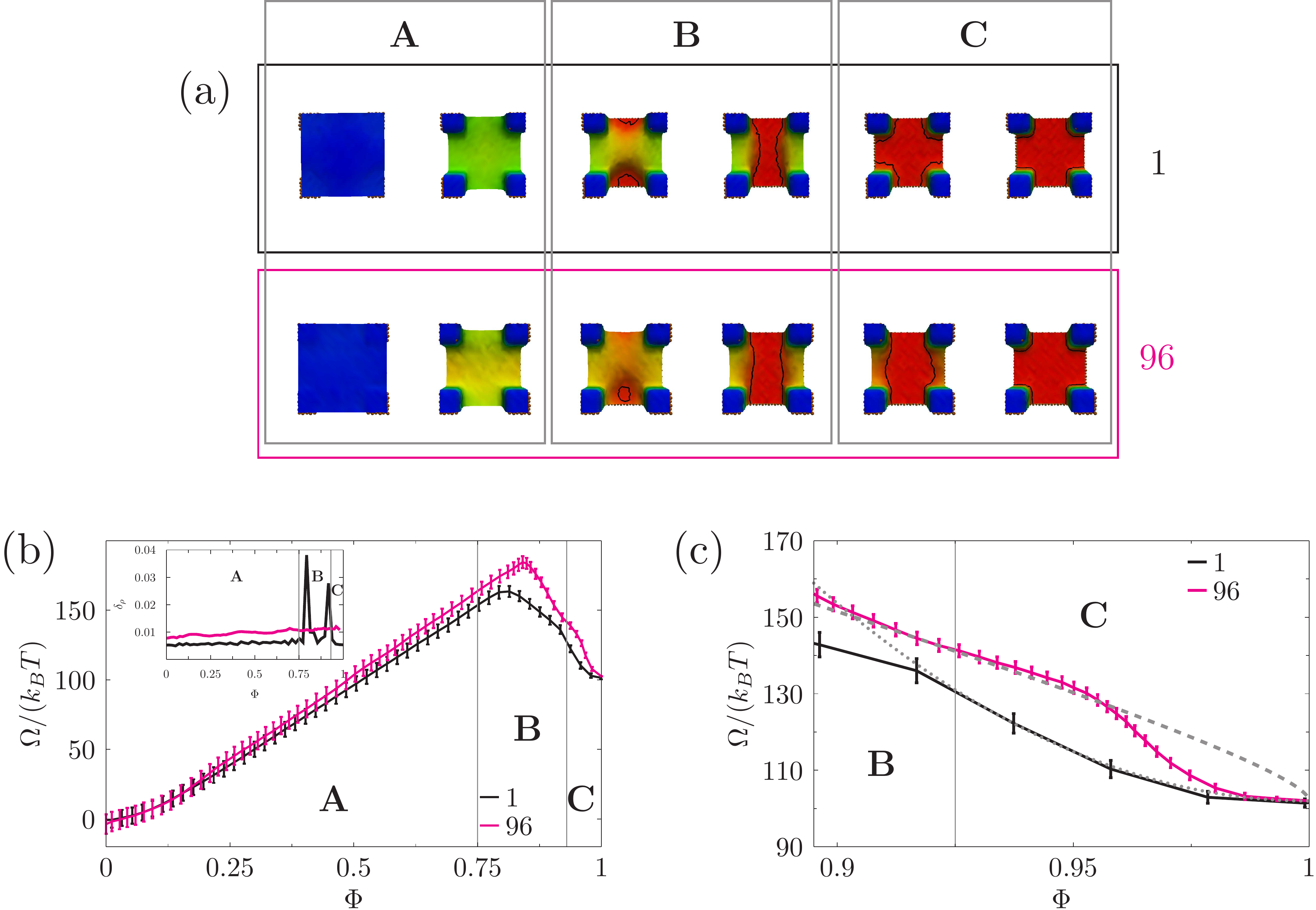}
        \caption{
a) Liquid meniscus along the Cassie-Wenzel transition for the $1\times 1$
system.  Black and violet frames correspond to $1$ and $96$ CVs, respectively.
The color code is the same as in Fig.  \ref{fig:percorso}: blue when the
meniscus is at the top of the pillar, red when it touches the bottom, and green
in between (see Supporting Information for Publication for the movies of
the wetting trajectories, files 1x1pillarCV96.gif, 1x1pillarCV1.gif, and
1x1regionC.gif). b) Free-energy profiles for the two sets of CVs. The inset
shows $\delta_{\rho}$ as a function of $\Phi$. c) Magnification of the region C
of the free-energy profile together with the  $c (1- \Phi)^2$ (grey dotted) and
$a (1 - \Phi)^{1/2}+b (1 - \Phi)$ (grey dashed) curves discussed in the text.
The coefficients $a$, $b$, and $c$ are obtained by fitting the two functions in
the domain C. To obtain a more careful comparison of the $1$ and $96$ CVs
profiles in this domain, the integration has been performed right to left,
which results in lower error bars.
\label{fig:profilipiccoli}} 
\end{figure*}

\subsubsection{Summary}

Let us close this section drawing some conclusions on the effect of the degree
of coarse graining on the study of the wetting/dewetting process.  Extreme
coarse-graining also leads to severe errors in the estimation of $\Delta
\Omega_{CW}$, up to $400~k_BT$; in addition, it also causes discontinuities in
the wetting path. Extreme coarse-graining affects the wetting barrier, which is
underestimated by $\approx 80~k_BT$ as compared to the finer density field
case. Overall, this suggests that quantitative predictions require a careful
choice of the CVs.  We remark that these effects are not due to the special
method used - RMD or string - analogous artifacts would be observed also with
US, BXD, TAMD, or other techniques.   A detailed analysis of string/RMD
\emph{vs} US and BXD is discussed in App.~\ref{app:RMDvsUSvsBXD}.

\subsection{$\bm 1 \bm \times \bm 1$ pillar system}
\label{sec:piccolo}

For the  $1 \times 1$  pillar system we perform an analysis similar to the one
of Sec.~\ref{sec:grande}. In this case we considered $1$ and $96$ CVs, having
the same grid spacing of the $9$ and $864$ CVs cases of the $3 \times 3$
pillars, respectively.  Fig. ~\ref{fig:profilipiccoli}  reports
$\rho(x,\Phi_i)$, $\Omega(\Phi_i)$, and $\delta_{\rho}(\Phi_i)$. The general
characteristics of the wetting mechanism observed for the $3 \times 3$ pillars
system are preserved also in the smaller sample: the liquid initially depins
from the pillars top, slides along their side with a flat meniscus (A), then
touches the bottom wall in a point between a pair of pillars (B), and finally
the vapor bubble is absorbed (C).  It is worth remarking that analogous
mechanisms have been reported in the literature for the case of water wetting a
pillared surface. In this case, simulations were performed using  INDUS using 1
CV analogous to the one used in the present work.~\cite{prakash2016}  The $96$
CVs case presents a smooth transition from one regime to another, with
$\delta_\rho$  remaining almost constant at $0.01$. On the contrary, with $1$
CV  $\delta_\rho$ shows abrupt jumps in correspondence of the morphological
transitions at the boundaries between domains A and B and B and C.  Videos of
the wetting process for the two cases are available in the Supporting Information.

The $1$ and $96$ CVs free energy profiles present no qualitative differences:
both curves are characterized by a single, well defined maximum in
correspondence of the configuration in which the meniscus touches the bottom
wall. Concerning quantitative aspects, while $\Delta \Omega_{CW}$ is the same
in both cases, the barrier of  $1$ CV is $20~k_BT$ lower than for $96$ CVs.
This is due to the entropic effect explained in the previous section. 

A more careful analysis of the free energy profile close to the Wenzel state
reveals differences in the curvature of the free energy as a function of the
filling fraction. In order to study such differences and to obtain a more
accurate free energy profile in region C, the path with $96$ CVs in the range
$\Phi \in [0.9,1]$ has been discretized in $32$ images (Fig.
~\ref{fig:profilipiccoli}(c)). With  $1$ CV a parabolic profile extending
over a broad range of filling fraction is observed, which is generally
attributed to Gaussian fluctuations of the liquid density as discussed in the
Lum-Chandler-Weeks theory of hydrophobicity. \cite{lum1999} A similar trend has
been observed by other authors using similar CVs .
\cite{remsing2015,amabili2016,prakash2016,giacomello2012} On the contrary, with
$96$ CVs the parabolic trend is observed over a much narrower domain, after
which the free energy profile shows a dependence of the type $a (1
-\Phi)^{1/2} + b (1- \Phi)$ typical of macroscopic theories, which accounts for
the liquid-vapor and solid-vapor interface energies~\cite{giacomello2012a} (see
Supporting Information for further details).  This is reflected on an
earlier formation of the liquid-vapor meniscus, as shown in the snapshots of
Fig.~\ref{fig:profilipiccoli}(a). The wetting mechanism identified with $1$ CVs
is characterized by a strong initial ($\approx 8~\%$) depletion of the density
of the liquid in the corrugation before the meniscus is formed. On the
contrary, the path obtained with $96$ CVs indicates a more modest $\approx
3-4~\%$ density depletion before the liquid-vapor interface
forms.\footnote{We remark that similar results are present also in the
$3 \times 3$ pillar case, in which a different slope of $\Omega$ in the domain
C is apparent (Fig.~\ref{fig:profili}).  The detailed analysis of the early
stage of dewetting, which requires a fine discretization of the path with a
large number of string images, was performed for the $1 \times 1$ case which is
computationally more convenient.} Considering that the $96$ CVs is a
superset of $1$ CV (overall density), i.e., the $96$ CV-space includes and
extends the $1$ CV one, the different behavior of the free energy indicates
that the overall density is inadequate for describing the wetting and dewetting
path. Since differences arise already at the beginning of dewetting,
relatively close to the Wenzel state, one must conclude that also in the
presence of moderate dewetting barriers, which occur closer to the fully
wet state, the prediction of paths and rates obtained from simulations based on
a single CV might be inaccurate.

In several works~\cite{remsing2015,prakash2016,amabili2016} the use of the
overall density as the only descriptor of the wetting/dewetting process has
been justified on the basis of the Lum-Chandler-Weeks theory~\cite{lum1999} of
hydrophobicity. However, the above observation that $1$ CV is insufficient to
characterize thermally activated wetting/dewetting processes suggests that the
extension of the theory of the hydrophobic effect out of its original
equilibrium scope may bring to overlooking important aspects of the transition.
Indeed, fluctuations of the overall density happen during dewetting but for the
process to take place they must be accompanied by other events, e.g., the
formation, bending, and displacement of the liquid meniscus.

\subsection{Size effect: comparison between $3 \times 3$ and $1 \times 1$ pillars systems}
\label{sec:confronto}

\begin{figure*}%[h!]
        \centering
        \includegraphics[width=0.7\textwidth]{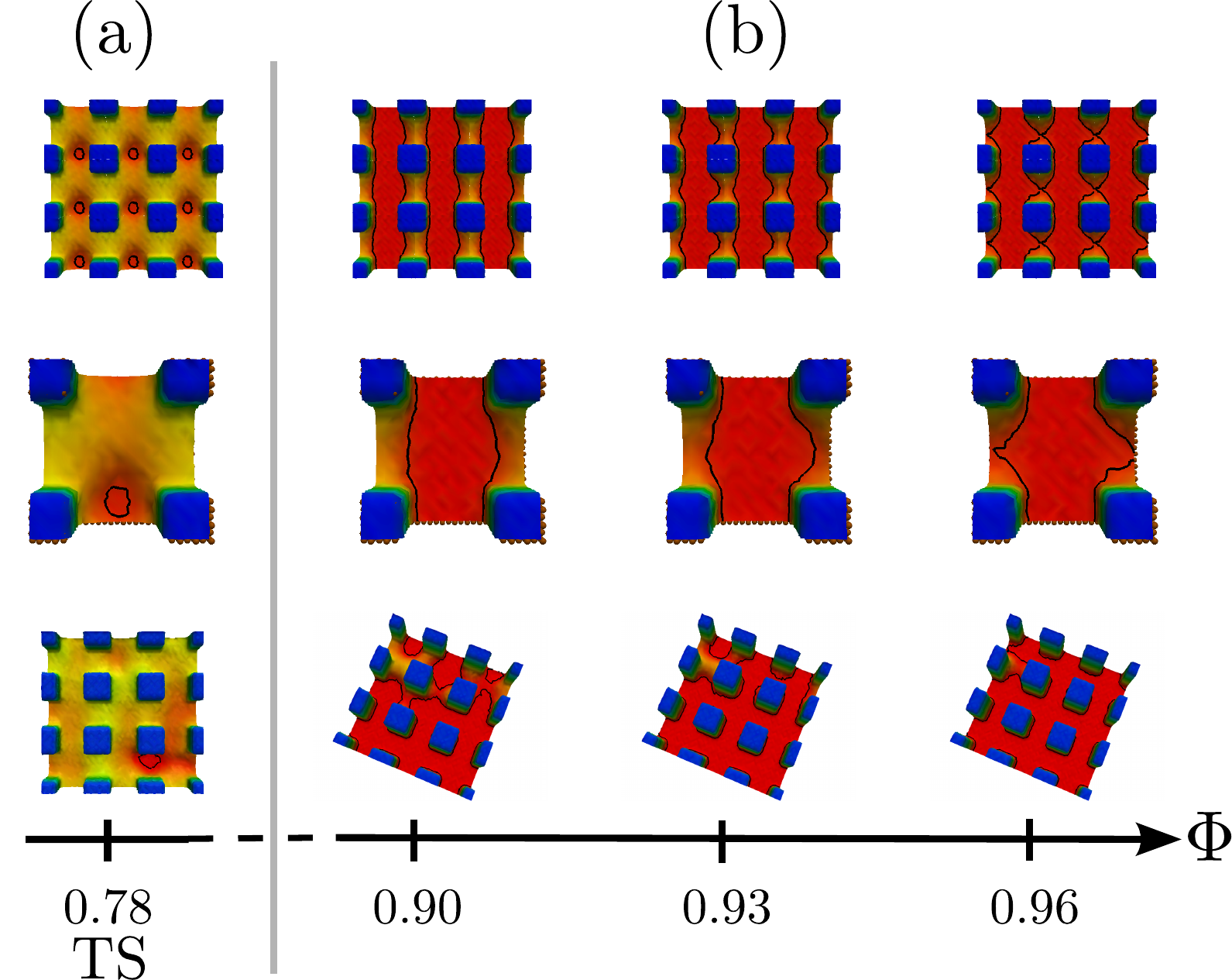}
				\caption{a) Meniscus at the transition states of the $1 \times
					1$ (center and top) and $3 \times 3$ (bottom) systems, for
					which  we considered $96$ and $864$ CVs, respectively, i.e.,
					they have the same grid spacing. The (black) three-phase
					contact line
					at the bottom wall
helps comparing the morphology of the menisci at the transition state.
This figure clearly illustrates that the $1 \times 1$ system is
insufficient for identifying possible breaks of the translational
symmetry of the meniscus. b) Menisci at selected values of liquid
fraction in the C domain. These snapshots show that the $1 \times 1$
system is insufficient for capturing the percolating network of vapor
bubbles characterizing this part of the wetting/dewetting path.}
        \label{fig:ts}
\end{figure*}

In the previous sections we focused on the effect of the choice of  CVs
on the wetting mechanism, energetics and kinetics; here we concentrate
on the effect of the size of the surface sample, the number of pillars
in the $x$ and $y$ directions. To this end we compare results obtained with
$96$ CVs for the  $1 \times 1$  surface and $864$ CVs for the  $3 \times
3$  one.
At the transition state the two systems present qualitatively similar
configurations. In both cases the liquid touches the bottom wall at a
point between two pillars (Fig.~\ref{fig:ts}). However, the $1 \times
1$ surface cannot reveal the complex nature of the final part of the
wetting. Indeed, for the $3 \times 3$ case we observe the formation of a percolating
(random) network of vapor bubbles, which is absorbed when the system
approaches the Wenzel state.~\cite{amabiliPRF} At variance with this
scenario, for the $1 \times 1$ surface, due to the constraints imposed
by the periodic boundary conditions, the network of vapor bubbles forms
between a pillar and its periodic image along one of the two
\emph{lattice} directions, $x$ and $y$ (compare top and bottom panels of
Fig.~\ref{fig:ts}).

\begin{figure}%[h!]
        \centering
        \includegraphics[width=0.45\textwidth]{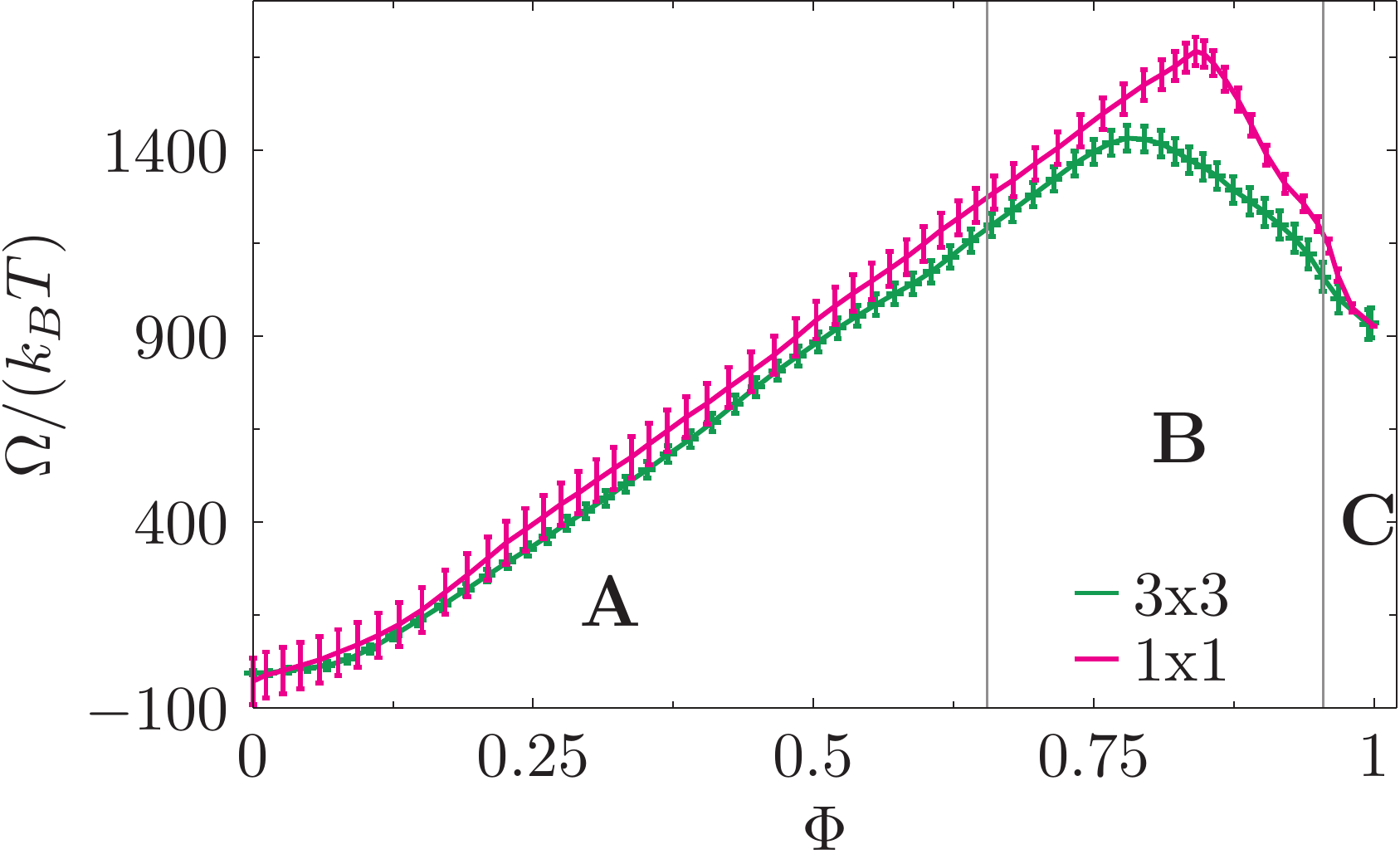}
	\caption{Free-energy profile for the
$1 \times 1$ (96 CV, violet line) and  $3 \times 3$ (864 CV, green line) systems.
The $1 \times 1$ profile  is obtained multiplying by $9$ the free-energy profile
of Fig.~\ref{fig:profilipiccoli}(b).}
        \label{fig:cfr_string}
\end{figure}

Concerning the quantitative comparison of results,
Fig.~\ref{fig:cfr_string} reports the free-energy
profiles for the $3 \times 3$ and $1 \times 1$ pillars systems. Given the
different size of the two systems, such an analysis is perfomed multiplying by
a factor $9$ the free-energy profile of Fig.\ref{fig:profilipiccoli}b. While,
as expected, the free-energy difference between Cassie and Wenzel states,
$\Delta\Omega_{CW}$, are in agreement, the barrier of the $1 \times 1$ pillar
system is $\approx 200~k_BT$ higher than the $3 \times 3$ one. Moreover, also
the position of the transition states is different in the two cases: $\Phi
\approx 0.75$ for the $3 \times 3$ pillars system as compared to $\Phi \approx
0.80$ for the $1 \times 1$ one.  This is the result of small size effect, more
specificallt, of the $9$ (periodic) repetition of the bent meniscus in the
replicated $1 \times 1$ system, see Fig.~\ref{fig:ts}.  $1 \times 1$ pillars
system. A fairer comparison can be obtained considering the free energy of the
transition state plus eight times that of the last flat meniscus configuration
of the  $1 \times 1$ system. However,  also in this case the $1 \times 1$
barrier is 180 $k_BT$ higher than the $3 \times 3$. This brings us to the
conclusion that the $1 \times 1$ system is affected by important finite-size
errors, due to the enforcement of an artificial symmetry.

%%%%%%%%%%%%%%%%%%%%%%%%%%%%%%%%%%%%%%%%%%%%%%%%%%%%%%%%%%%%%%%%%%%%%%%%%%%
\section{Conclusions \label{sec:conclusioni}}
In this work the Cassie-Wenzel transition has been studied using rare event
methods, namely the string and RMD techniques. The collective variable used is
the density field at various levels of coarse graining.  Results show the
importance of the choice of the collective variables in the case of wetting and
dewetting: extreme coarse graining of the density field, often used in the
recent literature, introduces qualitative and quantitative artifacts. These
artifacts include the  identification of a sequence of physically disconnected
density fields for the Cassie-Wenzel transition path as well as wrong estimates
of the relative energy between the Cassie and Wenzel states and the barrier
separating them.

We have also investigated the effect of the size of the sample representing the
surface, namely the number of pillars in the simulation box. Results show that
a single unit cell ($1 \times 1$ pillar system) is insufficient when dealing
with interconnected textures.  Indeed, the periodic boundary conditions impose
a non-physical translational symmetry of the meniscus leading to a incomplete
picture of the wetting mechanism. At least for pillars, in order to capture the
detailed wetting mechanism, the simulated system must be large enough to
capture the local deformation of the meniscus forming a \emph{liquid finger}
which touches the bottom wall between two pillars. Concerning dewetting, the
branch between the Wenzel and transition states  consists of a (random)
percolating network of vapor bubbles, which cannot be formed in the $1 \times
1$ pillar system. These simulation pitfalls preclude the use of too small
samples for deriving design criteria for preventing the wetting or improving
the superhydrophobicity recovery. 

The results of this work seem to indicate that the simplest simulations of
thermally activated wetting with few collective variables and small periodic
boxes should be taken \emph{cum grano salis}: the information they yield is
only qualitative and could possibly be used as a computationally inexpensive
way to identify trends or to perform parametric studies.\cite{prakash2016}
However, the detailed physics of the wetting/dewetting processes could be
qualitatively different from the simulated one and the value of the free-energy
barriers could be off by tens of $k_B T$.  Depending on the application and on
the goal of the work, one should choose what is the best compromise between a
physically sound picture of the phenomena and computationally convenient ways
to explore different cases in order to devise design strategies, e.g., for
preventing the complete wetting or helping the recovery of the Cassie state.
\emph{A posteriori}, one could evaluate the choice of CVs: typical symptoms of
the deficiency of the adopted description is the presence of non-differentiable
points in the free-energy profiles or the appearance of jumps in the density
along the wetting/dewetting processes.

\section{acknowledgement}

The research leading to these results has received funding from
the European Research Council under the European Union's
Seventh Framework Programme (FP7/2007-2013)/ERC Grant agreement n. [339446].  
We acknowledge PRACE for awarding us access to resource
FERMI based in Italy at Casalecchio di Reno.

\appendix

\section{The string algorithm}
\label{app:string}

The string algorithm aims at finding a path in the CV space $\left \{\bm
N(\alpha) \right \}_{\alpha=0,1}$ satisfying the condition that the force
orthogonal to the path, via a metric matrix $M$, is zero (Eq. \ref{eq:MFEP}):

\begin{equation}
\left [ \hat M(\bm N(\alpha)) \nabla_N \Omega(\bm N(\alpha)) \right ]_\perp = 0 
\label{eq:App-MFEP}
\end{equation}

In practice, the path $\left \{\bm N(\alpha) \right \}_{\alpha=0,1}$ is
discretized in $L$ images, and $\alpha$ takes values $\alpha = n/(L-1)$
with $n=0,\dots,L-1$. As already mentioned in the main text, the images
are enforced to be at a constant distance from each other (Eq.~\ref{eq:arclengthParam}).

One can introduce the following artificial dynamics
\begin{eqnarray}
{d N(\alpha,\tau) \over d\tau} = - \left [ \hat M(\bm N(\alpha, \tau)) \nabla_N \Omega(\bm N(\alpha, \tau)) \right ]_\perp
\label{eq.App-string}
\end{eqnarray}

\noindent where $\tau$ is an artificial time bearing no direct physical
meaning. For $\tau \rightarrow \infty$, $N(\alpha,\tau)$ converges to the
solution of Eq.~\ref{eq:App-MFEP}. The r.h.s. of Eq.~\ref{eq.App-string}
can be computed projecting out the component of the force parallel to the path
at the present $\tau$:
\begin{eqnarray}
\left [ \hat M(\bm N(\alpha, \tau)) \nabla_N \Omega(\bm N(\alpha, \tau)) \right ]_\perp &=& \\
\left (1 - \bm \xi \bm \xi^T \right)  \hat M(&\bm N&(\alpha, \tau)) \nabla_N \Omega(\bm N(\alpha, \tau)) \nonumber
\end{eqnarray}

\noindent with $\bm \xi$ tangent to the path and $\bm \xi^T$ its transpose. The
numerical determination of the tangent to a discrete path can be inaccurate,
which brings to numerical instabilities in the iterative procedure of
Eq.~\ref{eq.App-string}. Instead of removing the tangential component of the
force, $\bm N(\alpha, \tau)$ is first evolved according to the entire force,
then the equidistance among images is restored. \cite{e2007} This second step
is performed by moving the images along the polygonal curve interpolating
the path. This renormalization step, by removing the effect of the tangential
component of the force, is formally equivalent to applying the projection.

\section{Relation between the probability density function of different collective variables}
\label{app:1vs2CVs}
Direct comparison of the free energies of different sets of collective
variables is, in general, not possible. However, in the case of the sets of CVs
used in this work, and because of the steep profile of the free energy, this
comparison is, to a large extent, meaningful. The reason of this is explained
in the following. 

\begin{figure*}%[h!]
        \centering
        \includegraphics[width=1.0\textwidth]{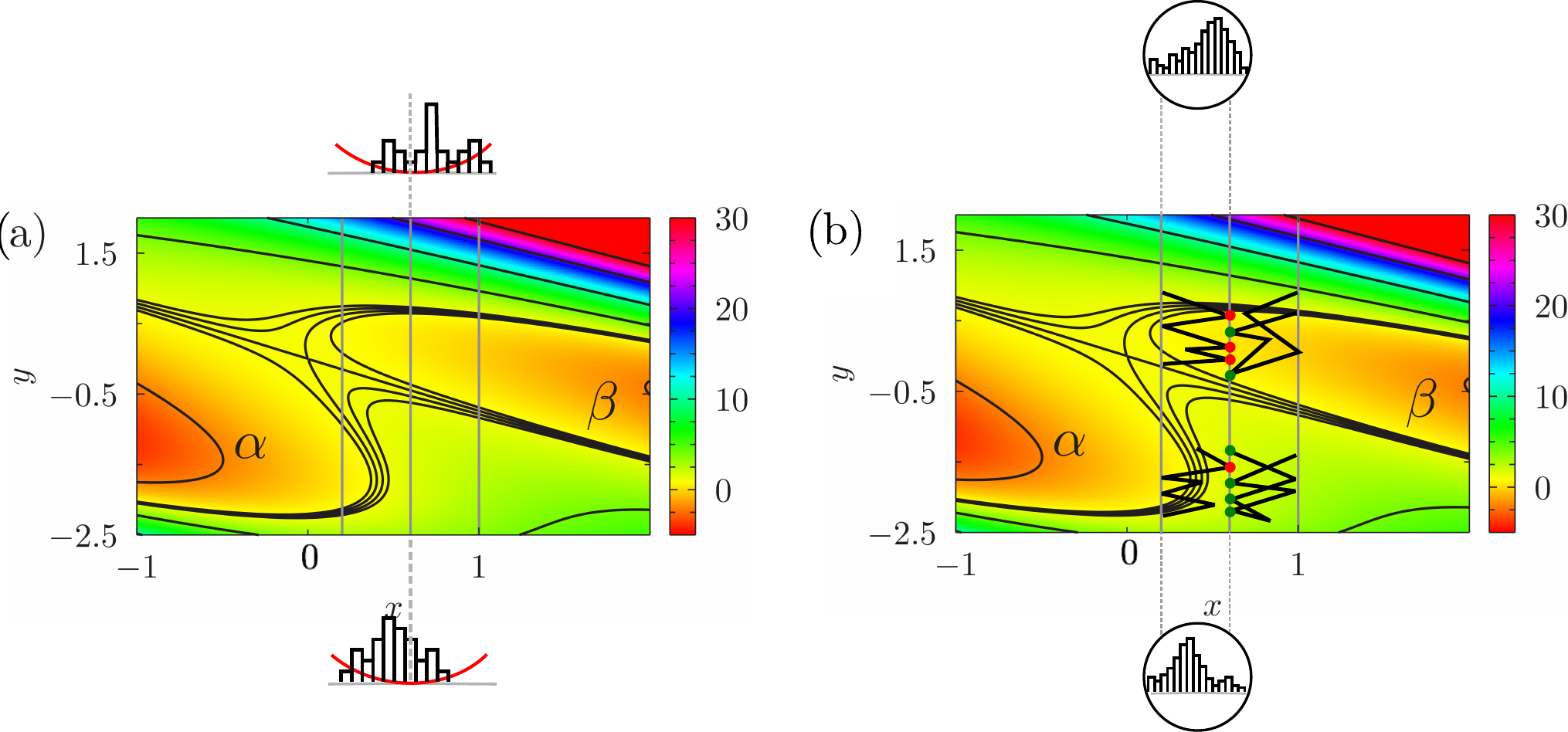}
        \caption{
Zoom of the $2$D polynomial potential of Fig.\ref{fig:toy} in the overlapping
region of basins $\alpha$ and $\beta$.  a) Results of US simulations: cartoon
of the probability density function of $x$ obtained from a thought US
simulation started from configurations belonging to $\alpha$ and $\beta$
attractive basins. b) Results of BXD simulations: cartoon of the probability
density function of $x$ within one box, corresponding trajectories and
boundaries hits (green and red circles). These examples illustrate that the
problems discussed in the main text are of general relevance and not limited to
RMD.
} 
\label{fig:confrotoMetodi}
\end{figure*}

Consider first the cases of $1$ and $2$ variables, the associated free energy $\Omega(x)$ and $\Omega(x, y)$  and probability density function $p(x) = \int dy\, p(x, y)$ and $p(x, y)$. To make the argument more straightforward, assume that the relevant ensemble is the canonical one; in this case  
\begin{equation}
\label{eq:App-1CVPDF}
p(x) = {\int dy\, \exp[-\beta \Omega(x, y)] \over \int dx\,dy\, \exp[-\beta \Omega(x, y)]}
\end{equation}

\noindent At low temperature, if the free energy has a minimum at each value of
$x$, $\Omega(x, y^0(x))$, the integrals in Eq.~\ref{eq:App-1CVPDF} can be
solved to a good approximation via the Laplace's method. This amounts to
replacing the integrand with the exponential of the second order expansion of
$\Omega(x, y)$ around $y^0(x)$.  Thus, $p(x) = p(x, y^0(x)) \sqrt{\pi / (\beta
\Omega''_y(x, y^0(x)))}$, with $\Omega''_y(x, y^0(x))$ the second derivative of
the free energy with respect to $y$ at the minimum.  The corresponding formula
for the free energy reads $\Omega(x) = \Omega(x, y^0(x)) -k_BT \log \sqrt{\pi /
(\beta \Omega''_y(x, y^0(x)))}$. 

The free energy is typically written with respect to a reference state,
say the state associated to the global minimum of the free energy
$x_\alpha$, and the relative free energy is $\Delta \Omega(x) =  \left (
\Omega(x, y^0(x)) - \Omega(x_\alpha, y^0(x_\alpha)) \right ) + k_BT \log
\Omega''_y(x, y^0(x)) / \Omega''_y(x_\alpha, y^0(x_\alpha))$. Henceforth
the relative free energy of the state is the same as the one of the
two-dimensional free energy at $(x, y^0(x))$ plus a term which is
proportional to the logarithm of the second derivative of the free
energy. This second term is expected to be exponentially negligible 
as compared to the first and goes to zero for $T \rightarrow 0$.

It is worth remarking that the above analysis holds true also in the
case in which $\Omega(x, y)$ presents more than one minimum in $y$ at a
given value of $x$. In this case, one could divide the integral at the
numerator of Eq.~\ref{eq:App-1CVPDF} in the sum of several integrals
centered around each minimum, and each integral can be solved with the
Laplace's method. The results is $p(x) = \sum_i p(x, y^{0,i}(x))
\sqrt{\pi / \left (\beta \Omega''_y(x, y^{0,i})\right )}$. Given the
exponential dependence of the probability density function on the free
energy, $p(x) \approx p(x, y^{0,\ast}(x))\, \sqrt{\pi / \left (\beta
\Omega''_y(x, y^{0,\ast})\right )}$, where $y^{0,\ast}(x)$ is the
absolute minimum at $x$. When two or more minima have the same
probability, $p(x) \approx p(x, y^{0,\ast}(x))\, \sum_i \sqrt{\pi / \left (\beta \Omega''_y(x, y^{0,i})\right )}$

In order for the above analysis to be useful in the present work, it must be
extended to the case in which the variable of which one wants to compute the
marginal probability is a function of the variables of the original set, not
just one element of the set.  For example, the $1$ CV variable set discussed in
the main text, consisting of the number of particles in the control volume, can
be written as the sum of the $864$ collective variables of the largest CVs set.
Consider first the transformation of variables $(N_1, \dots, N_{n_{CV}})
\rightarrow (N, N_2, \dots, N_{n_{CV}})$, with $N = \sum_{i=1,n_{CV}} N_i$. The
Jacobian of this transformation is $1$, which implies that $p(N, N_2, \dots,
N_{n_{CV}}) = p(N_1, \dots, N_{n_{CV}})$. One can then use the argument of the
previous paragraphs to show that $\Omega(N) \approx \Omega(N_1(N), \dots,
N_{n_{CV}}(N))$, where $(N_1(N), \dots, N_{n_{CV}}(N))$ is the configuration
identified by the string/RMD calculation at $N$. 

The above arguments prove that at low $T$ the free energy profiles with
all the CVs sets should be similar (provided that they are computed
along the same path).

\section{Comparison among string/restrained MD, umbrella sampling and boxed molecular dynamics}
\label{app:RMDvsUSvsBXD}

In this work we have analyzed the effect of the choice of the level of
coarse graining of the discretized density field on the mechanism and
energetics of the wetting process performing string and RMD
calculations. However, present results are not limited to the special
rare-event simulation technique adopted: analogous artifacts and
drawbacks affect simulations performed using other techniques with the
same CVs. The most used techniques for the study of wetting have been
umbrella sampling (US)\cite{TorrieValleau} and boxed MD (BXD);\cite{Glowacki:2009hqa} in this section we perform a
comparative analysis of these methods and explain why one should expect
analogous problems with these techniques when using the average density
in a given region as CV.

US relie on an algorithm which is similar to RMD: they both consist in
performing MD with an augmented potential of the form of $\tilde V(\bm r; \bm
N) = V(\bm r) + \lambda/2 \left (\phi(\bm r) - N \right )^2$. However, in US
the parameter $\lambda$ is, typically, set to much smaller values than in RMD.
Thus, the system is allowed to explore a larger space as compared to RMD, in
which the system visits only configurations consistent with the condition $\bm
\phi(\bm r) = \bm N$.  In US one first computes the free energy within the
range of collective variable values explored at the present value of $\bm N$.
Then, the branches of the free energy profiles of different umbrella potentials
are joined by imposing that they take the same value over the overlapping
regions of CVs. More elaborate and efficient techniques exist to compute the
free energy from US simulations~\cite{WHAM} but their discussion is beyond of
the scope of the present article.  In the presence of morphological
transitions, when a value of the CV in one or the other basin corresponds to
different configurations, the hypothesis of same probability density for same
value of the CV is not satisfied, which might bring to large errors in the
reconstructed free-energy profile (Fig.~\ref{fig:confrotoMetodi}(a)).

In BXD there is no augmented potential, and the atoms evolve according to the
physical potential $V(\bm r)$. However, the system is confined within a
\emph{box} in the (single) CV space, defined by a lower and upper value of the
collective variable. This box is equipped with reflecting boundary conditions,
which allows to extensively sample regions of overall low probability. The
calculation of the free energy profiles in BXD is a three-steps procedure;
first, one determines the free energy difference between neighboring boxes from
the rate of transition of forward and backward trajectories between them:
$\Delta \Omega_{n-1,n} = - k_BT \log(k_{n-1,n}/k_{n,n-1})$, with $k_{n-1,n}$
and $k_{n,n-1}$ rates of transition from box $n$ to box $n-1$ and \emph{vice
versa}, respectively. From the complete set of $\Delta \Omega_{n-1,n}$ for all
pairs of consecutive boxes one can compute the probability $p_n$ that the
system is within a  given box. Secondly, the probability density of the CV within
each box, $p_n(N)$, is computed from the histogram of the corresponding
trajectory. Thirdly, the Landau free energy is obtained from the logarithm of the
total probability $p(N) = p_n \times p_n(N)$, i.e., \emph{via}
Eq.~\ref{eq:Landau}.  When the system is characterized by two disjoint initial
and final basins as in Fig.~\ref{fig:confrotoMetodi}(b), the flux of
trajectories across a given interface is different if the system is in one or in
the other. Indeed, also $p_n(N)$ depends on which attractive basin the system
is in. Thus, also in this case, inadequate CVs may yield wrong free energy
profiles.
 
Summarizing, the error in the estimation of the free energy is not
associated to a specific rare-event  technique (RMD, US, BXD, etc.),
rather, it depends on the choice of CVs.

%%%%%%%%%%%%%%%%%%%%%%%%%%%%%%%%%%%%%%%%%%%%%%%%%%%%%%%%%%%%%%%%%%%%%%%%%%%
%%%%%%%%%%%%%%%%%%%%%%%%%%%%%%%%%%%%%%%%%%%%%%%%%%%%%%%%%%%%%%%%%%%%%%%%%%%
\bibliography{biblio}% Produces the bibliography via BibTeX.

\end{document}